\documentclass[12pt]{article}
\usepackage{graphicx}

\usepackage[british]{babel}
\usepackage[dvips]{geometry}
\usepackage{amsmath}
\usepackage{amsfonts}
\usepackage{amssymb}
\usepackage[mathscr]{eucal}
\usepackage{empheq}

\usepackage{epstopdf}
\usepackage{psfrag}
\usepackage{pstricks,pst-fill,pst-text,pst-node,pstricks-add}

\usepackage{hyperref}

\usepackage{tikz}
\usetikzlibrary{patterns,arrows,decorations.pathreplacing} 

\newcounter{braid}
\newcounter{strands}
\pgfkeyssetvalue{/tikz/braid height}{1cm}
\pgfkeyssetvalue{/tikz/braid width}{1cm}
\pgfkeyssetvalue{/tikz/braid start}{(0,0)}
\pgfkeyssetvalue{/tikz/braid colour}{black}
\pgfkeys{/tikz/strands/.code={\setcounter{strands}{#1}}}

\makeatletter
\def\cross{%
  \@ifnextchar^{\message{Got sup}\cross@sup}{\cross@sub}}

\def\cross@sup^#1_#2{\render@cross{#2}{#1}}

\def\cross@sub_#1{\@ifnextchar^{\cross@@sub{#1}}{\render@cross{#1}{1}}}

\def\cross@@sub#1^#2{\render@cross{#1}{#2}}

\def\render@cross#1#2{
  \def\strand{#1}
  \def\crossing{#2}
  \pgfmathsetmacro{\cross@y}{-\value{braid}*\braid@h}
  \pgfmathtruncatemacro{\nextstrand}{#1+1}
  \foreach \thread in {1,...,\value{strands}}
  {
    \pgfmathsetmacro{\strand@x}{\thread * \braid@w}
    \ifnum\thread=\strand
    \pgfmathsetmacro{\over@x}{\strand * \braid@w + .5*(1 - \crossing) * \braid@w}
    \pgfmathsetmacro{\under@x}{\strand * \braid@w + .5*(1 + \crossing) * \braid@w}
    \draw[braid] \pgfkeysvalueof{/tikz/braid start} +(\under@x pt,\cross@y pt) to[out=-90,in=90] +(\over@x pt,\cross@y pt -\braid@h);
    \draw[braid] \pgfkeysvalueof{/tikz/braid start} +(\over@x pt,\cross@y pt) to[out=-90,in=90] +(\under@x pt,\cross@y pt -\braid@h);
    \else
    \ifnum\thread=\nextstrand
    \else
     \draw[braid] \pgfkeysvalueof{/tikz/braid start} ++(\strand@x pt,\cross@y pt) -- ++(0,-\braid@h);
    \fi
   \fi
  }
  \stepcounter{braid}
}

\tikzset{braid/.style={double=\pgfkeysvalueof{/tikz/braid colour},double distance=1pt,line width=2pt,white}}

\newcommand{\braid}[2][]{%
  \begingroup
  \pgfkeys{/tikz/strands=2}
  \tikzset{#1}
  \pgfkeysgetvalue{/tikz/braid width}{\braid@w}
  \pgfkeysgetvalue{/tikz/braid height}{\braid@h}
  \setcounter{braid}{0}
  \let\g=\cross
  #2
  \endgroup
}
\makeatother

\newcommand{\brw}{17}

\usepackage{color}

\makeatletter
\@addtoreset{equation}{section}
\makeatother

\def\begeqar{\begin{eqnarray}}
\def\endeqar{\end{eqnarray}}
\def\begeq{\begin{equation}}
\def\endeq{\end{equation}}

\def\r#1{(\ref{#1})}

\def\wgta#1#2#3#4{\hbox{\rlap{\lower.35cm\hbox{$#1$}}
\hskip.2cm\rlap{\raise.25cm\hbox{$#2$}}
\rlap{\vrule width1.3cm height.4pt}
\hskip.55cm\rlap{\lower.6cm\hbox{\vrule width.4pt height1.2cm}}
\hskip.15cm
\rlap{\raise.25cm\hbox{$#3$}}\hskip.25cm\lower.35cm\hbox{$#4$}\hskip.6cm}}

\def\wgtb#1#2#3#4{\hbox{\rlap{\raise.25cm\hbox{$#2$}}
\hskip.2cm\rlap{\lower.35cm\hbox{$#1$}}
\rlap{\vrule width1.3cm height.4pt}
\hskip.55cm\rlap{\lower.6cm\hbox{\vrule width.4pt height1.2cm}}
\hskip.15cm
\rlap{\lower.35cm\hbox{$#4$}}\hskip.25cm\raise.25cm\hbox{$#3$}\hskip.6cm}}

\newcommand\be            {\begin{equation}}
\newcommand\ee            {\end{equation}}

\newcommand{\ATL}[1]{\mathsf{T}^a_{#1}}

\newcommand{\TL}[1]{TL_{#1}}

\newcommand{\q}{\mathfrak{q}}
\newcommand{\ffrac}[2]{\mbox{\footnotesize$\displaystyle\frac{#1}{#2}$}}
\newcommand{\half}{%
  \mathchoice{\ffrac{1}{2}}{\frac{1}{2}}{\frac{1}{2}}{\frac{1}{2}}}

\newcommand{\tensor}{\otimes}

\newcommand{\fus}{\times_{f}}
\newcommand{\fusV}{\times_{\rm CFT}}
\newcommand{\afusV}{\widehat{\times}_{\rm CFT}}

\newcommand{\aSt}{\mathcal{W}}

\newcommand{\StTL}[1]{\mathcal{W}_{#1}}

\newcommand{\StJTL}[2]{\mathcal{W}_{#1,#2}}

\newcommand{\Hom}{\mathrm{Hom}}
\newcommand{\Ind}[2]{\mathrm{Ind}_{#1}^{#2}}
\newcommand{\Res}[2]{\mathrm{Res}_{#1}^{#2}}
\newcommand{\Rep}{\mathrm{Rep}}

\newcommand{\oC}{\mathbb{C}}
\newcommand{\oZ}{\mathbb{Z}}

\newcommand{\afus}{\,\widehat{\times}_{\!f}\,}

\newcommand{\afusm}{\,\widehat{\times}^{-}_{\!f}\,}

\newcommand{\drawu}{
 	\draw[thick, dotted] (-0.05,0.5) arc (0:10:0 and -7.5);
 	\draw[thick, dotted] (-0.05,0.55) -- (3.25,0.55);
 	\draw[thick, dotted] (3.25,0.5) arc (0:10:0 and -7.5);
	\draw[thick, dotted] (-0.05,-0.85) -- (3.25,-0.85);
	\draw[thick] (0.3,0.5) arc (0:10:20 and -3.75);
	\draw[thick] (2.7,-0.81) arc (0:10.6:-30 and 3.75);

	\draw[thick] (0.9,0.5) arc (0:10:40 and -7.6);
	\draw[thick] (1.5,0.5) arc (0:10:40 and -7.6);
	\draw[thick] (2.1,0.5) arc (0:10:40 and -7.6);
	\draw[thick] (2.7,0.5) arc (0:10:40 and -7.6);
}

\newcommand{\AStTL}[2]{\aSt_{#1,#2}}
\newcommand{\bAStTL}[2]{\overline{\mathcal{W}}_{#1,#2}}
\newcommand{\VirN}{\boldsymbol{\mathcal{V}}}
\newcommand{\Verma}[1]{\mathsf{V}_{#1}}
\newcommand{\bVerma}[1]{\overline{\mathsf{V}}_{#1}}
\newcommand{\IrrV}[1]{\mathsf{X}_{#1}}
\newcommand{\IrrVb}[1]{\overline{\mathsf{X}}_{#1}}

\newcommand{\ipic}[3][-0.5]{\raisebox{#1\height}{\scalebox{#3}{\includegraphics{#2.pdf}}}}

\newcommand{\tensore}{\boxtimes}

\begin{document}
\thispagestyle{empty}

\begin{center}
\Large{A fusion for the periodic  Temperley--Lieb algebra \\ and its continuum limit }

\vskip 1cm

{\large Azat M.\ Gainutdinov $^{1,2,3}$, Jesper Lykke Jacobsen $^{4,5,6}$ and Hubert Saleur $^{6,7}$}

\vspace{1.0cm}
{\sl\small $^1$  Laboratoire de Math\'ematiques et Physique Th\'eorique CNRS/UMR 7350,
 F\'ed\'eration Denis Poisson FR2964,
Universit\'e de Tours,
Parc de Grammont, 37200 Tours, 
France\\}
{\sl\small $^2$ Institut f\"ur Mathematik,
Mathematisch-naturwissenschaftliche Fakult\"at,
Universit\"at Z\"urich, Winterthurerstr.\,190, CH-8057 Z\"urich, Switzerland\\}
{\sl\small $^3$ Fachbereich Mathematik, Universit\"at Hamburg, Bundesstr.\,55, 20146 Hamburg, Germany\\}
{\sl\small $^4$ Laboratoire de physique th\'eorique, D\'epartement de physique de l'ENS,
\'Ecole Normale Sup\'erieure,
UPMC Univ.~Paris 06, CNRS, PSL Research University, 75005 Paris, France\\}
{\sl\small $^5$ Sorbonne Universit\'es, UPMC Univ.~Paris 06, \'Ecole Normale
Sup\'erieure,
CNRS, Laboratoire de Physique Th\'eorique (LPT ENS), 75005 Paris, France\\}
{\sl\small $^6$  Institut de Physique Th\'eorique, CEA Saclay,
Gif Sur Yvette, 91191, France\\}
{\sl\small $^7$ Department of Physics and Astronomy,
University of Southern California,
Los Angeles, CA 90089, USA\\}

\end{center}

\begin{abstract}

The equivalent of fusion in boundary conformal field theory (CFT) can be realized quite simply in the context of lattice models by essentially glueing two open spin chains. This has led to many developments, in particular in the context of chiral logarithmic CFT. 

We consider in this paper a possible  generalization of  the idea to the case of bulk conformal field theory. This is of course considerably more difficult, 
since there is no obvious way of merging two closed spin chains into a big one.
 In an earlier paper, two of us had proposed a ``topological'' way of performing this operation in the case of models based on the affine Temperley-Lieb  (ATL) algebra, by exploiting the associated braid group representation and skein relations. In the present work, we establish---using, in particular, Frobenius reciprocity---the resulting fusion rules for standard modules of ATL in the generic as well as partially degenerate cases. These fusion rules have a simple interpretation in the continuum limit.
 However, unlike in the chiral case this interpretation does not match the usual fusion in non-chiral CFTs. 
  Rather, it corresponds to the glueing of the right moving component of one conformal field with the left moving component of the other. 
\end{abstract}

\newpage
\setcounter{tocdepth}{2}
\tableofcontents

\section{Introduction}

The study of relations between lattice models and conformal field theories (CFT) has a long history, and stems from several different motivations. Most recently, it has been part of a push to understand better Logarithmic CFTs \cite{LCFTreview}, which are seemingly too hard to be tackled from the top down, by abstract CFT constructions. Instead, it has proven extremely valuable to study in detail the corresponding lattice regularizations, and to infer from these results about the continuum limit. 

Crucial in this approach has been  the construction of a lattice equivalent of conformal fusion in the chiral case. 
The idea can be expressed simply in terms of spin chains \cite{ReadSaleur07-2}---i.e., multiple tensor products of vector spaces such as the  fundamental representations of $sl(2)$ or $sl(2|1)$---where a certain algebra, such as the Temperley-Lieb (TL) algebra, acts and commutes with the symmetry. Two spin chains can then be ``glued'' by taking modules of the algebra for each of them, adding a generator that connects the chains, and defining a ``lattice" fusion of these modules using induction. 
In formulas this reads (see below for details)
\begin{equation}
\label{eq:TLfus-Ind-open-intro}
M_1\fus M_2 = \Ind{\TL{N_1}\tensor\TL{N_2}}{\TL{N}} M_1\tensor M_2 \ .
\end{equation}
Remarkably, this fusion product can be put in one-to-one correspondence~\cite{GV} with fusion of chiral fields in CFT, including in the non-unitary cases where indecomposable and logarithmic modules appear~\cite{GaberdielKausch}. 

Underlying the success of this approach is the deep and still mysterious connection between  the TL algebra (and its ``blob algebra'' generalizations) on the one hand, and the Virasoro algebra on the other hand~\cite{GJSV}.%
\footnote{Similar connections are known for more complicated lattice algebras such as the Birman-Wenzl-Murakami algebra and more complicated chiral algebras such as the $N=1$ super Virasoro algebra \cite{PRT13}.}
A similar connection is known to exist between the product of left and right Virasoro algebras and the affine Temperley-Lieb algebra (ATL), which is the natural generalization of the TL  algebra for periodic spin chains. For recent works on this topic, see \cite{GRS1,GRS2,GRS3,GJRSV,GRSV1,ZW17}. This raises the interesting question of whether an analog of non-chiral fusion in CFT can be defined in the context of  the ATL algebra.

The reasons why spin chains related with the product of left and right Virasoro algebras must be periodic is that CFTs in radial quantization \cite{DfMS} have a Hamiltonian acting on an ``equal radial time'' circle, which becomes the periodic chain after discretization.%
\footnote{This is in contrast with the case of a chiral Virasoro algebra, which is naturally associated with a boundary CFT, and thus a Hamiltonian acting on a half circle in radial quantization.}
The periodicity  makes it very difficult to define fusion: by analogy with what happens in the chiral case, one would like to be able to bring two chains close to each other and connect them with an extra interaction. But periodicity seems to require the chains to be cut open first, and it is not clear how to do this in a natural, ``generic'' way. 

Recently, however, two of us \cite{GS} proposed a way to solve this problem, and delineated the basic principles to define the analog of (\ref{eq:TLfus-Ind-open-intro}) in the case of the ATL algebra. The purpose of the present work is to calculate explicitly  the fusion of standard modules of ATL, and to study, using the relationship with product of left and right Virasoro algebras \cite{GJRSV}, the  corresponding non-chiral fusion in the associated CFTs. 

Our paper is organized as follows. We start in Section~\ref{sec:fusion} by recalling the definition of fusion for ATL modules and important results from \cite{GS}. We also recall salient features of the ATL algebra, including the definition and classification of standard modules from~\cite{GL}. In Section~\ref{sec:Frobenius} we discuss Frobenius reciprocity, which is a crucial tool that we use later to calculate ATL fusion rules. In Section~\ref{sec:generic}
we explain our general strategy to calculate fusion rules, and give our main result in~(\ref{genfusii}). We then discuss the principal properties of these fusion rules, including   associativity, braiding and stability with respect to increasing number of cites. In Section~\ref{sec:partdeg} we extend our results to the ``partially degenerate case'', where the deformation parameter $\q$ is still generic, but the momentum-related parameter in ATL standard modules---denoted $z$ here---is not. In this paper, we however do not analyze cases of $\q$ a root of unity, which is reflected by the fact that on the CFT side in the continuum limit we consider generic central charges only.
In Section~\ref{sec:CFTlimit} we discuss the interpretation of our fusion results in the context of CFTs:
we explain how the fusion of ATL modules corresponds, in the continuum limit, to a glueing of the left-moving sector of one field with the right-moving sector of the other. Unfortunately, this glueing is not what one would normally call fusion in non-chiral CFT.
In Section~\ref{sec:Conclusion} we gather conclusions, and discuss possible generalizations and extensions of this work.
Some technical details about fusion are discussed in two appendices. Appendix~\ref{sec:appA} gathers the explicit branching rules of generic ATL standard modules on $N=6$ sites with respect to the standard modules of two embedded periodic subalgebras of any size $N_1$ and $N_2$ sites (with $N=N_1+N_2$). Appendix \ref{sec:appB} discusses further the issue of uniqueness of the  embedding that underlies our fusion construction.

\medskip

We give here a brief list of our notations, consistent with the works \cite{GRS1,GRS2,GRS3,GJRSV,GRSV1}:
\begin{itemize}

\item$\TL{N}(m)$ --- the (ordinary) Temperley--Lieb algebra on $N$ sites with loop fugacity $m$,

\item$\ATL{N}(m)$ --- the affine Temperley--Lieb algebra,

\item$\StTL{j}$ --- the standard modules over $\TL{N}(m)$, 

\item$\AStTL{j}{z}$ --- the standard modules over $\ATL{N}(m)$,

\item$\VirN$  --- the product of left and right Virasoro algebras,

\item $\Verma{r,s}$ --- Verma modules with the conformal weight $h_{r,s}$ where $r$ and $s$ are the Kac labels,

\item $\IrrV{r,s}$ --- irreducible Virasoro (Kac) modules of the conformal dimension $h_{r,s}$ in the case of generic central charge,

\item$\tensore$ --- outer product of left and right Virasoro representations,

\item $\IrrV{r,s}\tensore\IrrVb{r',s'}$ --- irreducible $\VirN$-modules,

\item$K_{r,s}(q)$  --- characters of  $\IrrV{r,s}$ with the formal variable $q$,
  
 \item $F_{j,z}$  ---  character or $(L_0,\bar{L}_0)$-graded dimension of $\StJTL{j}{z}$ in the continuum limit,

\item $\fus$ --- lattice fusion in the open case, i.e.\ for ordinary TL algebra,

\item $\fusV$ --- continuum limit of the fusion un the open case,

\item$\afus$ --- lattice fusion in the periodic case, i.e.\ for the affine TL algebra,

\item$\afusm$ --- other possible lattice fusion for ATL, obtained by exchanging ``above'' and ``under'',

\item$\afusV$ --- continuum limit of the fusion in the periodic case.

\end{itemize}

\section{Fusion in the periodic case}\label{sec:fusion}

\subsection{Fusion}\label{sec:fusion-ATL}

The affine Temperley--Lieb (aTL) algebra $\ATL{N}(m)$ is an associative algebra over $\oC$ generated by $u$, $u^{-1}$ , and $e_j$, with $j\in \oZ/N\oZ$, satisfying  the defining relations
\begin{eqnarray}
e_j^2&=&me_j,\nonumber\\
e_je_{j\pm 1}e_j&=&e_j,\label{TL}\\
e_je_k&=&e_ke_j\qquad(j\neq k,~k\pm 1),\nonumber
\end{eqnarray}
which are the standard Temperley-Lieb relations but defined for the indices modulo $N$, and
\begin{eqnarray}
ue_ju^{-1}&=&e_{j+1},\nonumber\\
u^2e_{N-1}&=&e_1\ldots e_{N-1}, \label{TLpdef-u2}
\end{eqnarray}
where  the indices $j=1, \ldots, N$ are  again interpreted modulo $N$.

$\ATL{N}(m)$  admits a representation in terms of  diagrams
on an annulus,
with  $N$ labeled sites on the inner and $N$ on
the outer rim.
The sites are connected in pairs,
and only configurations that can be represented using non-crossing simple curves inside the
annulus are allowed. Curves connecting sites on opposite rims are called {\em through-lines}.
Diagrams related by an isotopy
leaving the labeled sites fixed are considered equivalent. We call such
(equivalence classes of) diagrams  \textit{affine} diagrams.  
 Multiplication $a\cdot b$ of two affine diagrams $a$ and $b$ is defined in a natural
way, by gluing $a$ around $b$. By this we mean precisely 
joining the inner rim of $a$ to the outer rim of the annulus of $b$, and
removing the sites at the junction, so that the outer rim of $a$ and the inner rim of $b$
becomes the rims of the product $a \cdot b$.
Whenever a closed contractible loop is
produced by this gluing, this loop must be
replaced by a numerical factor~$m$, that we often parametrize by $\q$ as $m=\q+\q^{-1}$. 
 Below we will use a slightly different graphical presentation: the diagrams are drawn
as rectangles with periodic boundary conditions where the inner rim of the annulus is mapped to the bottom of the rectangle.

We note that the  diagrams
in this algebra allow winding of through-lines around the annulus {\sl any
integer number of times}, and different windings result in independent
algebra elements. Moreover, in the ideal of zero through-lines, any number of
non-contractible loops can pile up inside the annulus. The algebra $\ATL{N}(m)$ is thus infinite-dimensional.

 \begin{figure}
\begin{center}
 \ipic{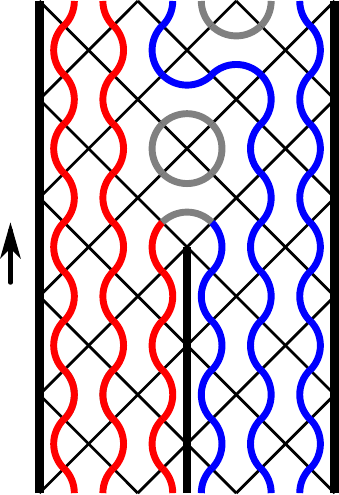}{1.2} 
 \put(-75,90){ $\TL{2(j_1+j_2)}$}
\put(-130,-5){ $\tau$}
\put(-95,-95){ $\TL{\color{red}2j_1}\,\tensor\,\TL{\color{blue}2j_2}$}
\end{center}
  \caption{Physical interpretation of the fusion of two standard TL modules $\StTL{j_1}[N_1]$ and $\StTL{j_2}[N_2]$ (in the picture, 
  $N_1=2j_1$ and $N_2=2j_2$ so that both standard modules are one-dimensional). On the lattice, the induction procedure can be seen as an event in imaginary
  time $\tau$, consisting in ``gluing'' the two standard modules by acting with an additional TL generator. }
  \label{FigFusion}
\end{figure}

The affine Temperley--Lieb  algebra is the natural object underlying many statistical mechanics models on a cylinder, just like the ordinary Temperley-Lieb algebra is the object describing these systems on strips, corresponding to open boundary conditions. In this latter case, it is easy to imagine a process of ``fusion'' of two systems on $N_1$ and $N_2$ sites, simply by putting them next to each other so as to form a system of $N=N_1+N_2$ sites, and adding an extra generator that connects the two subsystems. This process is illustrated diagrammatically in Figure~\ref{FigFusion}. Note that in statistical mechanics, the physical systems have often the form of ``spin chains'', that is, the Temperley-Lieb generators act on a space which is the tensor product of certain (super) Lie algebra or quantum group representations on every site $j$.  We will sometimes use this language of chains, especially in the section devoted to the continuum limit.

In more mathematical terms, we can take $M_1$ and $M_2$ to be two modules over $\TL{N_1}$ and $\TL{N_2}$ respectively.
The tensor product $M_1\tensor M_2$ is then a module over
 the product $\TL{N_1}\tensor\TL{N_2}$ of the two algebras.
Using the standard  embedding, we consider this
 product of algebras as a subalgebra in $\TL{N}$. The fusion of the two modules 
  is then defined as the  module induced from this subalgebra, {\it i.e.}
\begin{equation}
\label{eq:TLfus-Ind-open}
M_1\fus M_2 = \Ind{\TL{N_1}\tensor\TL{N_2}}{\TL{N}} M_1\tensor M_2 \ ,
\end{equation}
where $\Ind{B}{A} M$ stands for the  representation induced from the $B$-module $M$, for a given subalgebra $B$ in $A$.
Practically, the induced modules can be computed as the so-called balanced tensor product $A\tensor_B M$, which is the quotient of $A\tensor M$ by imposing the relations $ab\tensor m = a\tensor bm$ for all $a\in A$, $b\in B$ and $m\in M$.  In our case it is then
\begin{equation}\label{fusfunc-TL-def}
M_1\fus M_2 = \TL{N}\tensor_{\bigl(\TL{N_1}\tensor\TL{N_2}\bigr)}\big(M_1\tensor M_2\big)\ .
\end{equation}

\bigskip

It is of course not so easy to do the same thing in the periodic case, since---unlike in the open case---there
is no such thing as a ``last site'' of the first subsystem, nor a ``first site'' of the second subsystem, across
which the two subsystems can be naturally glued. Rather, we need a prescription for smoothly merging
the two cylinders of circumference $N_1$ and $N_2$, each representing a subsystem, into a bigger cylinder
of circumference $N=N_1+N_2$. A strategy for doing this was first proposed in \cite{GS} and will be stated
precisely below. In intuitive terms it amounts to imposing the periodic boundary condition within each
subsystem by bypassing---using a knot-theory inspired construct known as braiding---the sites of the other
subsystem. This merging of cylinders is traditionally represented by a diagram resembling a pair of
trousers (with the legs representing the subsystems), but we chose not to do so here, as we feel that such
a diagram over-simplifies the underlying idea. Instead we shall represent the braidings diagramatically below.

To give an example of the construction in~\cite{GS}, the  two translations operators arising from  $\ATL{N_1}$ and  $\ATL{N_2}$ are now represented, after fusion, by 
(with $N_1=3$ and $N_2=2$, and we distinguish two subsystems by different colors, used only here)
%
\begin{equation}\label{u1-def}
 \begin{tikzpicture}
 \node[font=\large]  at (-1.5,-2.2) {\mbox{} $u^{(1)}\;\;\mapsto$ \mbox{}\qquad};
\drawu 
\braid[braid colour=black,strands=5,braid width=\brw,braid start={(-0.3,-0.88)}]  {\g_4^{-1}\g_3^{-1}}
\node[font=\large]  at (4.0,-2.2) {\mbox{}\qquad \(=\)};
\end{tikzpicture}
\qquad
 \begin{tikzpicture}
 	\draw[thick, dotted] (-0.05,0.5) arc (0:10:0 and -7.5);
 	\draw[thick, dotted] (-0.05,0.55) -- (3.25,0.55);
 	\draw[thick, dotted] (3.25,0.5) arc (0:10:0 and -7.5);
	\draw[thick, dotted] (-0.05,-0.85) -- (3.25,-0.85);

	\draw[red,thick] (0.3,0.5) arc (0:10:20 and -3.75);
	\draw[red,thick] (0.9,0.5) arc (0:10:40 and -7.6);
	\draw[red,thick] (1.5,0.5) arc (0:10:40 and -7.6);
	

	\draw[red,thick] (1.6,-0.81) .. controls (1.7,-0.3)   .. (2.1,-0.25);
	\draw[red,thick] (2.1,-0.25) -- (3,-0.25);
	\draw[blue,thick] (2.1,0.5) -- (2.1,-0.15);
	\draw[blue,thick] (2.1,-0.35) -- (2.1,-0.8);
	\draw[blue,thick] (2.7,0.5) -- (2.7,-0.15);
	\draw[blue,thick] (2.7,-0.35) -- (2.7,-0.8);
	\draw[red,thick] (3,-0.25) .. controls (3.13,-0.2)   .. (3.21,-0.1);
\end{tikzpicture}
\end{equation}
and 
\begin{equation}\label{u2-def}
 \begin{tikzpicture}
 \node[font=\large]  at (-1.5,-0.2) {\mbox{} $u^{(2)}\;\;\mapsto$ \mbox{}\qquad};
 
\drawu 

\braid[braid colour=black,strands=5,braid width=\brw,braid start={(-0.28,3.6)}]  { \g_3 \g_2 \g_1}

\node[font=\large]  at (4.0,-0.2) {\mbox{}\qquad \(=\)};
\end{tikzpicture}
\qquad
 \begin{tikzpicture}
 	\draw[thick, dotted] (-0.05,0.5) arc (0:10:0 and -7.5);
 	\draw[thick, dotted] (-0.05,0.55) -- (3.25,0.55);
 	\draw[thick, dotted] (3.25,0.5) arc (0:10:0 and -7.5);
	\draw[thick, dotted] (-0.05,-0.85) -- (3.25,-0.85);
	\draw[blue,thick] (0.0,-0.18) .. controls (0.15,0.08)   .. (0.43,0.1);
	\draw[blue,thick] (0.58,0.1) -- (1.02,0.1);
	\draw[blue,thick] (1.18,0.1) -- (1.62,0.1);
	\draw[blue,thick] (1.77,0.1) .. controls (2.1,0.2)   .. (2.3,0.5);

	\draw[blue,thick] (2.9,-0.81) arc (0:10:-20 and 3.75);

	\draw[red,thick] (0.5,0.5) -- (0.5,-0.8);
	\draw[red,thick] (1.1,0.5) --  (1.1,-0.8);
	\draw[red,thick] (1.7,0.5) -- (1.7,-0.8);
	\draw[blue,thick] (2.75,0.5) arc (0:10:40 and -7.6);

\end{tikzpicture}
\end{equation}
where one should resolve each braiding by the ``skein"  relations
\begin{equation}
 g_j = {\rm i} \left( \q^{1/2} - \q^{-1/2} e_j \right) \,, \qquad
 g_j^{-1} = -{\rm i} \left( \q^{-1/2} - \q^{1/2} e_j \right) \,,
 \label{skein-rel}
\end{equation}
where the two braid operators, $g_j$ and $g_j^{-1}$, and the Temperley-Lieb generator $e_j$,
can be represented graphically by
\begin{equation}\label{eq:g-diagr}
g_j =
\begin{tikzpicture}[baseline={([yshift=-12pt]current bounding box.north)}]
 \draw[thick] (0,0.6)--(0.6,0);
 \draw[thick] (0,0)--(0.25,0.25);
 \draw[thick] (0.35,0.35) -- (0.6,0.6);
\end{tikzpicture} \ \,, \qquad g_j^{-1} = 
\begin{tikzpicture}[baseline={([yshift=-12pt]current bounding box.north)}]
 \draw[thick] (0,0.6)--(0.25,0.35);
 \draw[thick] (0.35,0.25)--(0.6,0);
 \draw[thick] (0,0)--(0.6,0.6);
\end{tikzpicture} \ \,, \qquad e_j =
\begin{tikzpicture}[baseline={([yshift=-12pt]current bounding box.north)}]
 \draw[thick] (0,0.6) arc (180:360:0.3 and 0.2);
 \draw[thick] (0.6,0.0) arc (0:180:0.3 and 0.2);
\end{tikzpicture} \ \,.
\end{equation}
Note that $g_j$ and $g_j^{-1}$ are indeed each other's inverse, due to the relation
$e_j^2 = (\q+\q^{-1}) e_j$.
The two translation operators corresponding to the two periodic subsystems, represented graphically in (\ref{u1-def})--(\ref{u2-def}), can be expressed algebraically in terms of the translation $u$ and the braid operators $g_j^{\pm 1}$ as
\begin{equation}
 u^{(1)} = u\, g_{N-1}^{-1} \cdots g_{N_1}^{-1} \,, \qquad
 u^{(2)} = g_{N_1} \cdots g_1\,u \,.
 \label{trans-op-subsystems}
\end{equation}
It is important to realize that the over/under structure of the two translation operators $u^{(1)}$ and $u^{(2)}$ is crucial to ensure the commutation $[u^{(1)},u^{(2)}]=0$.

We note that there is a certain degree of arbitrariness in this way of proceeding. In particular, one could as well define fusion by swapping over- and underpasses, resulting in the diagrams:

\begin{equation}\label{utilde1-def}
 \begin{tikzpicture}
 \node[font=\large]  at (-1.5,-2.2) {\mbox{} $\tilde{u}^{(1)}\;\;\mapsto$ \mbox{}\qquad};
\drawu 
\braid[braid colour=black,strands=5,braid width=\brw,braid start={(-0.3,-0.88)}]  {\g_4\g_3}
\node[font=\large]  at (4.0,-2.2) {\mbox{}\qquad \(=\)};
\end{tikzpicture}
\qquad
 \begin{tikzpicture}
 	\draw[thick, dotted] (-0.05,0.5) arc (0:10:0 and -7.5);
 	\draw[thick, dotted] (-0.05,0.55) -- (3.25,0.55);
 	\draw[thick, dotted] (3.25,0.5) arc (0:10:0 and -7.5);
	\draw[thick, dotted] (-0.05,-0.85) -- (3.25,-0.85);

	\draw[thick] (0.3,0.5) arc (0:10:20 and -3.75);
	\draw[thick] (0.9,0.5) arc (0:10:40 and -7.6);
	\draw[thick] (1.5,0.5) arc (0:10:40 and -7.6);
	

	\draw[thick] (1.6,-0.81) .. controls (1.7,-0.3)  .. (2.,-0.25 );
		\draw[thick] (2.2,-0.25) -- (2.6,-0.25);
	\draw[thick] (2.1,0.5) -- (2.1,-0.8);
	\draw[thick] (2.7,0.5) -- (2.7,-0.8);
		\draw[thick] (2.8,-0.25) .. controls (3.13,-0.2)   .. (3.21,-0.1);
\end{tikzpicture}
\end{equation}
and 
\begin{equation}\label{utilde2-def}
 \begin{tikzpicture}
 \node[font=\large]  at (-1.5,-0.2) {\mbox{} $\tilde{u}^{(2)}\;\;\mapsto$ \mbox{}\qquad};
 
\drawu 

\braid[braid colour=black,strands=5,braid width=\brw,braid start={(-0.28,3.6)}]  { \g_3^{-1} \g_2^{-1} \g_1^{-1}}

\node[font=\large]  at (4.0,-0.2) {\mbox{}\qquad \(=\)};
\end{tikzpicture}
\qquad
 \begin{tikzpicture}
 	\draw[thick, dotted] (-0.05,0.5) arc (0:10:0 and -7.5);
 	\draw[thick, dotted] (-0.05,0.55) -- (3.25,0.55);
 	\draw[thick, dotted] (3.25,0.5) arc (0:10:0 and -7.5);
	\draw[thick, dotted] (-0.05,-0.85) -- (3.25,-0.85);
	\draw[thick] (0.0,-0.18) .. controls (0.15,0.08)   .. (0.58,0.1);
	\draw[thick] (0.58,0.1) -- (1.02,0.1);
	\draw[thick] (1.02,0.1) -- (1.77,0.1);
	\draw[thick] (1.77,0.1) .. controls (2.1,0.2)   .. (2.3,0.5);

	\draw[thick] (2.9,-0.81) arc (0:10:-20 and 3.75);

	\draw[thick] (0.5,0.5) -- (0.5,0.17);
	\draw[thick] (0.5,0.) -- (0.5,-0.8);
	\draw[thick] (1.1,0.5) --  (1.1,0.2);
	\draw[thick] (1.1,0.) --  (1.1,-0.8);
	\draw[thick] (1.7,0.5) -- (1.7,0.2);
	\draw[thick] (1.7,0.) -- (1.7,-0.8);

	\draw[thick] (2.75,0.5) arc (0:10:40 and -7.6);

\end{tikzpicture}
\end{equation}
In the sequel we shall however stick to the conventions set out by (\ref{u1-def})--(\ref{u2-def}).

The ``closing'' Temperley-Lieb generators for the fused systems are represented similarly by (for the same example) 
\begin{equation}\label{e0-1-def}
 \begin{tikzpicture}
\node[font=\large]  at (-1.1,0.4) {\mbox{} $e_0^{(1)} =$};
\draw[thick] (0,0.5) -- (0.3,0.5) arc (270:360:0.2) -- (0.5,1.5);
\draw[thick] (0,0.2) -- (0.3,0.2) arc (90:0:0.2) -- (0.5,-0.8);
\draw[thick] (1.0,1.5) -- (1.0,-0.8);
\draw[thick] (3,0.5) -- (1.7,0.5) arc (270:180:0.2) -- (1.5,1.5);
\draw[thick] (3,0.2) -- (1.7,0.2) arc (90:180:0.2) -- (1.5,-0.8);
\draw[thick] (2.0,1.5) -- (2.0,0.58);
\draw[thick] (2.0,0.42) -- (2.0,0.28);
\draw[thick] (2.0,0.12) -- (2.0,-0.8);
\draw[thick] (2.5,1.5) -- (2.5,0.58);
\draw[thick] (2.5,0.42) -- (2.5,0.28);
\draw[thick] (2.5,0.12) -- (2.5,-0.8);
\end{tikzpicture}
\end{equation}

\begin{equation}\label{e0-2-def}
 \begin{tikzpicture}
\node[font=\large]  at (-1.1,0.4) {\mbox{} $e_0^{(2)} =$};
\draw[thick] (1.58,0.5) -- (1.8,0.5) arc (270:360:0.2) -- (2.0,1.5);
\draw[thick] (1.58,0.2) -- (1.8,0.2) arc (90:0:0.2) -- (2.0,-0.8);
\draw[thick] (0.5,1.5) -- (0.5,-0.8);
\draw[thick] (1.0,1.5) -- (1.0,-0.8);
\draw[thick] (1.5,1.5) -- (1.5,-0.8);
\draw[thick] (3,0.5) -- (2.7,0.5) arc (270:180:0.2) -- (2.5,1.5);
\draw[thick] (3,0.2) -- (2.7,0.2) arc (90:180:0.2) -- (2.5,-0.8);
\draw[thick] (0,0.5) -- (0.42,0.5);
\draw[thick] (0.58,0.5) -- (0.92,0.5);
\draw[thick] (1.08,0.5) -- (1.42,0.5);
\draw[thick] (0,0.2) -- (0.42,0.2);
\draw[thick] (0.58,0.2) -- (0.92,0.2);
\draw[thick] (1.08,0.2) -- (1.42,0.2);
\end{tikzpicture}
\end{equation}
The corresponding general algebraic expressions are
\begin{eqnarray}
 e_0^{(1)} &=& g_{N_1} \cdots g_{N-1} e_0 g_{N-1}^{-1} \cdots g_{N_1}^{-1} \,, \nonumber \\
 e_0^{(2)} &=& g_0^{-1} \cdots g_{N_1-1}^{-1} e_{N_1} g_{N_1-1} \cdots g_0 \,.
 \label{e01-e02-alg}
\end{eqnarray}

On the formal level, the formulas~\eqref{u1-def}, \eqref{u2-def} and~\eqref{e01-e02-alg} define an embedding of the two small periodic systems, or
more precisely the corresponding algebras $\ATL{N_1}$ and $\ATL{N_2}$, into the big
one $\ATL{N=N_1+N_2}$.   We will denote
the generators of the two periodic subalgebras as $e_j^{(1)}$ and $e_k^{(2)}$ respectively. Under this
embedding, the ordinary (open) TL generators are mapped as
\begin{eqnarray}
 e_j^{(1)} &\mapsto& e_j \,, \quad \, \, \; \mbox{ for } j=1,\ldots,N_1-1 \,, \nonumber \\
 e_k^{(2)} &\mapsto& e_{N_1+k} \,, \mbox{ for } k=1,\ldots,N_2-1 \,.
 \label{TL-gens-embedding}
\end{eqnarray}

This completes our review of the construction of  the subalgebras $\ATL{N_1}$ and $\ATL{N_2}$ inside $\ATL{N}$. It was checked in~\cite{GS} that all the defining relations of the subalgebras
are indeed satisfied. The fusion of any two representations of $\ATL{N_1}$ and $\ATL{N_2}$ is then formally defined as the induction from the subalgebra $\ATL{N_1}\tensor\ATL{N_2}$, similarly to what was done  in the open case:
\begin{equation}\label{eq:TLfus-Ind}
M_1\afus M_2 = \Ind{\ATL{N_1}\tensor\ATL{N_2}}{\ATL{N}} M_1\tensor M_2 \ ,
\end{equation}
where we recall that
$\Ind{B}{A} M := A\tensor_B M$ stands for the  representation induced from the $B$-module $M$, for a given subalgebra $B$ in $A$. 
According to the discussion below~\eqref{eq:TLfus-Ind-open}
 the induced module can be computed as the  balanced tensor product
\begin{equation}\label{fusfunc-TL-def}
M_1\afus M_2 = \ATL{N}\tensor_{\bigl(\ATL{N_1}\tensor\ATL{N_2}\bigr)}\big(M_1\tensor M_2\big)\ .
\end{equation}
Examples of the direct calculation using this formula are given in~\cite[Sec.\ 4.3]{GS}.

\subsection{Standard modules}\label{sec:St-mod}

We next need to recall some result about modules.
The standard modules $\StJTL{j}{z}[N]$  over $\ATL{N}(m)$, which are generically irreducible, are parametrized by pairs $(j,z)$, with a half-integer $j$ and a non-zero complex number $z$
whose meanings we now explain.

By convention, the evolution operator (transfer matrix or Hamiltonian) of the system acts on the outer rim of the annulus. If we impose a fixed number $2j$ of through-lines (satisfying $0\leq 2j \leq N$, and with $2j$ having the same parity as $N$), the arcs that connect the inner rim to itself are immaterial for the action of the evolution operator and we can hence ignore them. It is thus sufficient to consider {\em uneven diagrams} with $N$ points on the outer rim and only the $2j$ end points of the through-lines on the inner rim; the latter $2j$ points are then called {\em free sites}.  
For $j>0$, the algebra action can cyclically permute the free sites, that is to say, rotate the inner rim of the annulus with respect to the outer rim. These rotations give rise to a
pseudomomentum, parametrized by $z$, whose precise definition will be given below. Once the effect of rotations has been taken into account via $z$, the inner rim of the annulus becomes
irrelevant, provided that we record the positions of the through-lines on the outer rim. The result is a so-called {\em link diagram} that represents the $N$ sites on the outer rim, of which
$2j$ through-line positions are shown as {\em strings} and the remaining $N-2j$ points are pairwise connected via non-crossing {\em arcs}.
For example, 
  \begin{tikzpicture}
 	\draw[thick, dotted] (-0.05,0.1) arc (0:10:0 and -3.5);
 	\draw[thick, dotted] (-0.05,0.1) -- (1.05,0.1);
 	\draw[thick, dotted] (1.05,0.1) arc (0:10:0 and -3.5);
	\draw[thick, dotted] (-0.05,-0.55) -- (1.05,-0.55);
	\draw[thick] (0,-0.2) arc (-90:0:0.25 and 0.25);
	\draw[thick] (0.4,0.05) arc (0:10:0 and -2.6);
	\draw[thick] (0.6,0.05) arc (0:10:0 and -2.6);
	\draw[thick] (1.0,-0.2) arc (-90:0:-0.25 and 0.25);
	\end{tikzpicture}
  \, and \, 
  \begin{tikzpicture}
 	\draw[thick, dotted] (-0.095,0.1) arc (0:10:0 and -3.5);
 	\draw[thick, dotted] (-0.095,0.1) -- (1.025,0.1);
 	\draw[thick, dotted] (1.0,0.1) arc (0:10:0 and -3.5);
	\draw[thick, dotted] (-0.095,-0.55) -- (1.025,-0.55);
  \draw[thick] (0,0.05) arc (0:10:0 and -2.6);
	\draw[thick] (0.2,0.05) arc (-180:0:0.25 and 0.25);
	\draw[thick] (0.9,0.05) arc (0:10:0 and -2.6);
	\end{tikzpicture}
are two possible link patterns corresponding to $N=4$ and $j=1$. By convention we identify the left and right sides of the framing rectangles, so the link patterns live in the annulus. These
link patterns form a basis of $\StJTL{j}{z}[N]$. The algebra $\ATL{N}(m)$ acts on the link patterns with the same graphical rules as for the diagrams in the annulus (in particular, giving a
weight $m$ to each closed loop), except that we have now imposed a fixed number $2j$ of strings, so the action of any generator in $\ATL{N}(m)$ that contracts two strings is set to zero.

In the original diagrammatic formulation, when $j\neq 0$, through-lines can wind around the annulus an arbitrary number of times. The module $\StJTL{j}{z}[N]$ is obtained by unwinding these lines at the price of numerical factor that will fix the pseudomomentum $z$. More precisely, whenever $2j$ through-lines wind counterclockwise around
the annulus $l$ times, we unwind them at the price of a factor $z^{2jl}$; similarly, for clockwise winding, the phase 
is $z^{-2jl}$~\cite{MartinSaleur,MartinSaleur1}. Alternatively, in the link pattern formulation, when a string traverses the side of the framing rectangle going towards the right (resp.\ left), it picks up the phase $z$ (resp.\ $z^{-1}$). 

These conventions give rise to a generically
irreducible module, by which we mean that the $\StJTL{j}{z}[N]$ thus defined is irreducible~\cite{GL} for generic values of $z$ and $m$.
Parametrizing as usual $m=\q+\q^{-1}$, we say that $m$ is generic if $\q$ is not a root of unity, while $z$ is generic if it is not an integer power of $\q$.

When $j=0$ (note this is only possible when $N$ is even), we can obtain any number of non-contractible loops. The module $\StJTL{0}{z}[N]$ is obtained by replacing each of these loops by the factor $z+z^{-1}$. This is compatible with the simple convention that any strand (whether of the string or arc type) that traverses the side of the framing rectangle acquires the phase $z^{\pm 1}$---note in particular that this leaves unchanged the weight of a contractible loop, since it makes an equal number of left and right traversals.

The dimensions of  $\StJTL{j}{z}[N]$ are found by elementary combinatorics to be given by 
 \begin{equation}\label{dim-dj}
 \hat{d}_{j}[N]\equiv \dim \StJTL{j}{z}[N] =
 \binom{N}{{N\over 2}+j},\qquad j\geq 0.
 \end{equation}
Note that these dimensions do not depend on $z$ (but modules with
different $z$ are not isomorphic). 

For generic values of $\q$, the modules $\StJTL{j}{z}[N]$ can be decomposed into a direct sum of standard modules $\StTL{j}[N]$ for the ordinary algebra $\TL{N}(m)$. These well-known modules are parametrized by the number $2j$ of through lines, and their dimension is 
\begin{equation}\label{dj}
d_j[N]=\left(\begin{array}{c} N\\{N\over 2}+j\end{array}\right)-\left(\begin{array}{c} N\\{N\over 2}+j+1\end{array}\right).
\end{equation}
We have then the simple result of the restriction to the TL subalgebra:
$$
\StJTL{j}{z}[N] = \bigoplus_{k=j}^{N/2} \StTL{k}[N] \ .
$$ 

\section{Frobenius reciprocity}\label{sec:Frobenius}

\subsection{The general result}
Following our discussion of the fusion---in terms of the gluing of two strips (resp.\ cylinders) in the open
(resp.\ periodic) case, see (\ref{eq:TLfus-Ind-open}) and (\ref{eq:TLfus-Ind})---the remaining question is now how to compute the induced modules.
In the remainder of this subsection we shall discuss the open case, but the
statements for the periodic case are identical, provided we replace $\TL{N}$ by $\ATL{N}$.

This mathematical formulation of the fusion as an induction has some advantages. It is first of all a well-posed computational problem and in many cases the result of the fusion---the so-called \textit{fusion rules}---can be computed directly and quite explicitly~\cite{GV,RSA,BRSA}, especially in the open case. However, in the periodic case it is still a very difficult task, and even for a very small number of sites the computation is quite involved~\cite[Sec.\,4.3]{GS}. But we can use instead a classical result in mathematics called ``Frobenius reciprocity'' (see \cite{Mackey}) 
 to relate the problem of induction with another problem, namely the restriction of representations from the big algebra on $N_1+N_2$ sites to the product of the two ``small'' TL subalgebras on $N_1$ and $N_2$ sites. The latter is technically much simpler to compute. To be more precise, computing the fusion rules means deducing certain positive integer numbers $N_{i,j}^k$:
\begin{equation*}
M_i\fus M_j =  \bigoplus_k N_{i,j}^k P_k
\end{equation*}
together with determining the indecomposable modules $P_k$ that arise when we fuse a given $M_i$ with $M_j$, like the standard modules introduced above. At least in the generic cases,%
\footnote{As already mentioned above, the open case is called generic if $\q$ in $m=\q+\q^{-1}$ is not a root of unity. In the periodic case we require in addition that $z$ is not an integer power of $\q$.} where the representation theory is semi-simple, this amounts to computing  dimensions of the spaces of linear maps 
from $M_i\fus M_j$ to $P_k$  
 that commute with  the $\TL{N}$ action.%
\footnote{In the non-semisimple cases (i.e., when $\q$ is a root of unity) one would need first to construct projective covers of irreducibles and study  $\mathrm{Hom}$ spaces of maps from the projective covers to $M_i\fus M_j$.}
By the Schur lemma these dimensions agree with the numbers $N_{i,j}^k$:
\be\label{eq:NijkHom}
N_{i,j}^k = \dim \Hom_{\TL{N}} (M_i\fus M_j, P_k) \ .
\ee

It is now important to note that for the $\Hom$ vector space in~\eqref{eq:NijkHom} we have an isomorphism with another space of linear maps but now commuting with the $\TL{N_1}\tensor\TL{N_2}$ action:
\begin{equation}\label{eq:Frob-TL}
\Hom_{\TL{N}} (M_i\fus M_j,P_k)  \cong  \Hom_{\TL{N_1}\tensor\TL{N_2}} \bigl(M_i\tensor M_j, P_k\big|_{\TL{N_1}\tensor\TL{N_2}}\bigr)\ .
\end{equation}
This is an instance of the so-called \textit{Frobenius reciprocity} which states in general  that for a given associative algebra $A$, its subalgebra $B\subset A$  and an $A$-module $V$ and a $B$-module $W$ there is an isomorphism
\begin{equation}\label{eq:Frob-gen}
\psi\colon \quad \Hom_{A} ( \Ind{B}{A}W,V) \xrightarrow{\;\sim\;}   \Hom_{B} (W,\Res{B}{A} V)\ ,
\end{equation}
where $\Res{B}{A}V:= V\big|_B$ is the restriction of $V$ to the subalgebra $B$.%
\footnote{In more abstract terms, the Frobenius reciprocity can be rephrased as follows: the induction functor $\Ind{}{}\colon \Rep B \to \Rep A$ is the {\sl left adjoint} to the restriction functor $\Res{}{}\colon \Rep A \to \Rep B$.}
Using~\eqref{eq:TLfus-Ind} and the Frobenius reciprocity~\eqref{eq:Frob-gen} for the case $A=\TL{N}$ and $B=\TL{N_1}\tensor\TL{N_2}$ considered as the subalgebra in~$A$, and setting $V=P_k$ and $W=M_i\tensor M_j$, we indeed have an
isomorphism in~\eqref{eq:Frob-TL}.

\subsection{Two rules}\label{sec:2rules}
Let us rephrase the formula~\eqref{eq:Frob-TL} in words as a  rule:
\begin{itemize}
\item[(R1)] \textit{a module $P_k$ appears in the fusion $M_i\fus M_j$ as many times as the restriction of $P_k$ to the ``small'' algebra contains the ordinary tensor product $M_i\tensor M_j$}.
\end{itemize}
Strictly speaking, this rule can be applied in the semi-simple case only, i.e.\ for generic values of~$\q$. In the non-semisimple case, the formula~\eqref{eq:Frob-TL} provides only dimensions of spaces of intertwining operators, and it requires a more detailed analysis in order to determine if a given $P_k$ is actually a direct summand in $M_i\fus M_j$. Also, it is clear from the rule (R1) that one has to test it against all possible indecomposable representations $P_k$ -- this is not a problem in the generic case when such a classification is known, however a classification is much more involved in degenerate cases, especially for the affine TL algebra discussed below.
\medskip

In the periodic case, we have a similar formulation of the fusion $\afus$ of the $\ATL{N}$ modules in terms of the induction~\eqref{eq:TLfus-Ind}.
 We have thus  an isomorphism of the spaces of linear maps respecting  $\ATL{N}$ and $\ATL{N_1}\tensor\ATL{N_2}$ actions:
\begin{equation}\label{eq:Frob-ATL}
\Hom_{\ATL{N}} \bigl(M_1\afus M_2, P_k\bigr)  \cong  \Hom_{\ATL{N_1}\tensor\ATL{N_2}} \bigl(M_1\tensor M_2, P_k\big|_{\ATL{N_1}\tensor\ATL{N_2}}\bigr)\ ,
\end{equation}
and therefore the same rule (R1) formulated above applies here as well.

\newcommand{\C}{\mathsf{C}}
\newcommand{\Irr}{\mathrm{Irr}}
\medskip

In conclusion, we reduced studying the fusion of (arbitrary) modules $M_i$ and $M_j$ in the open and periodic cases to computing the restrictions, which is a much simpler task. 
Furthermore, this restriction can be practically studied using spectral properties of some ``nice" operators.
Since this is going to be our main tool below, we explain the procedure in some detail.

Assume that we have a central element $\C$ in the ``big'' algebra $A_N$ (here $A$ stands for $\TL{}$ or~$\ATL{}$) that satisfies the non-degeneracy condition
\be
{\rm C1}\colon \quad \text{the spectrum of $\C$ on $\Irr$ is non-degenerate},
\ee
where by $\Irr$ we denote the set of all irreducible representations of $A$. We will denote the unique eigenvalue of $\C$ on a given irreducible $V\in\Irr$ by $\C_V$. The condition C1 then assumes that $\C_V=\C_W$ iff $V$ is isomorphic to $W$. For the two subalgebras $A_{N_1}$ and $A_{N_2}$ in $A_N$ we also have the corresponding elements $\C^{(1)}$ and $\C^{(2)}$. Then the restriction of a given $A_N$-module $M$ onto the subalgebra $A_{N_1}\tensor A_{N_2}$ can be studied by analysing the spectrum of the two operators $\C^{(1)}$ and $\C^{(2)}$ on $M$. We thus can formulate the second and more practical rule:
\begin{itemize}
\item[(R2)]
\textit{a module $P_k$ appears in the fusion $M_i\fus M_j$ 
if and only if the following conditions are satisfied:
\begin{enumerate}
\item  the spectrum of  $\C^{(1)}$ on $P_k$  contains $\C_{M_i}$ and the multiplicity is divisible by $\dim M_i \times \dim M_j$\ ,
\item   the spectrum of  $\C^{(2)}$ on $P_k$  contains $\C_{M_j}$ and the multiplicity is divisible by $\dim M_i \times \dim M_j$\ ,
\item   the spectrum of  $\C^{(1)}\C^{(2)}$ on $P_k$  contains $\C_{M_i}\C_{M_j}$ and the multiplicity is divisible by $\dim M_i \times \dim M_j$\ .
\end{enumerate}
}
\end{itemize}
 The same also applies for the affine TL fusion $\afus$. We note once again that the rule (R2)
 allows to determine a decomposition of the restricted module only in the generic cases. Otherwise the spectrum gives only a partial information on the decomposition.
 
We give below several explicit examples of the computation using a slight modification of the rule (R2), as one usually does not have such a nice central element $\C$ but rather a family of elements that essentially meets the above conditions on the spectrum. To demonstrate the idea underlying the calculations we start with the open case.

\subsection{The open case}
\label{sec:frob-fuse-open}
We now consider the application of the general result~\eqref{eq:Frob-TL} to the calculation of the open TL fusion rules. For generic $\q$, the algebra $\TL{N}$ is semisimple and thus our rule (R1) applies here, and to compute the multiplicity of $\StTL{j}$ in $\StTL{j_1}\fus\StTL{j_2}$ it is enough to compute the multiplicity of $\StTL{j_1}[N_1]\tensor\StTL{j_2}[N_2]$ in $\StTL{j}[N_1+N_2]$ considered as the representation of $\TL{N_1}\tensor\TL{N_2}$. The latter multiplicity we find using the graphical realization of $\StTL{j}$'s.

Let us  consider first the case $N_1=N$ and $N_2=1$. We are then looking for the restriction of the standard modules $\StTL{j}[N+1]$ to the subalgebra 
$\TL{N}\tensor\TL{1}$ where $\TL{1}$ is the trivial one-dimensional algebra, so consisting of only the unit element. In this case, there are only two types of configurations: one where the rightmost site is a string, and the other where an arc connects the $k$th site with the $(N+1)$th one. The link states of the first type obviously form an invariant subspace under the action of $\TL{N}\tensor\TL{1}$ and give a basis in the module $\StTL{j_1}[N]\tensor\StTL{j_2}[1]$ with $j_2=\half$ and $j_1 = j - \half$ (since the left subsystem then has one less string). Taking the quotient by this submodule, we are left with the configurations of the second type only. In the quotient space, these are also invariant under the action, and since the connecting arc roles as a string in either subsystem (due to the restriction to $\TL{N_1}\tensor\TL{N_2}$) we obtain $j_1 = j + \half$ and $j_2 = \half$ in that case. In total we have
\be
 \StTL{j}[N+1] \big|_{\TL{N}\tensor\TL{1}} = \StTL{j-\half}[N]\tensor\StTL{\half}[1] \oplus \StTL{j+\half}[N]\tensor\StTL{\half}[1]\ .
\ee
Applying then our rule (R1) we obtain the well-known result~\cite{ReadSaleur07-2,GV} for TL fusion:
\be
\StTL{j}[N]\fus\StTL{\half}[1] = \StTL{j-\half}[N+1] \oplus \StTL{j+\half}[N+1]\ .
\ee

This type of argument can be generalised, and amounts effectively to having any number of connecting arcs between the two subsystems, when viewed as a restriction of the standard module $\StTL{j}[N]$ with $N=N_1+N_2$ to the subalgebra $\TL{N_1}\tensor\TL{N_2}$. The final result is
\begin{equation}
\StTL{j}[N]\big|_{\TL{N_1}\tensor\TL{N_2}} = \bigoplus^*_{j_1,j_2} \StTL{j_1}[N_1]\otimes \StTL{j_2}[N_2] \,,
\label{fuse-open-TL}
\end{equation}
where the asterisk indicates that the sum is over $j_1, j_2 \geq 0$, such that $2j_1$ (resp.\ $2j_2$) has the same parity as $N_1$ (resp.\ $N_2$), and with the constraint $|j_1-j_2|\leq j\leq j_1+j_2$. To wit, note that  the upper bound on $j$ is explained by observing that the number of strings in the whole system, $2j$, can at
most be equal to the sum of the number of strings in either subsystem, $2(j_1+j_2)$. Indeed this corresponds
to a situation with no arc connecting the two subsystems. The lower bound on~$j$ corresponds similarly
to replacing each of the strings in one subsystem---namely the subsystem having the least number strings---by
an arc connecting it to a string in the other subsystem.  The number of unpaired strings remaining in the other
subsystem is then $2|j_1-j_2|$, explaining the minimal value of $j$ for the whole system. 
Finally, on such subsystems for each choice of $(j_1,j_2)$ within the bounds just described, the action of two algebras $\TL{N_1}$ and $\TL{N_2}$ is well-defined, of course modulo states with fewer number of through-lines, and this action clearly agrees with the corresponding $\StTL{j_1}[N_1]$ and  $\StTL{j_2}[N_2]$. 

We then observe, that the result~\eqref{fuse-open-TL} leads to the general fusion, here we again follow (R1):
\be
\StTL{j_1}[N_1]\fus\StTL{j_2}[N_2] =  \bigoplus^*_{j} \StTL{j}[N]
\ee
where the asterisk sum-condition corresponds to the bounds established above: $|j_1-j_2|\leq j\leq j_1+j_2$.
This result of TL fusion was also established  before~\cite{ReadSaleur07-2,GV} using other means.

Alternatively, one could follow the prescription in (R2) using the central element $\C$ in TL algebra whose diagrammatic form is the following
\begin{equation}
\C = \begin{tikzpicture}[baseline={([yshift=-25pt]current bounding box.north)},scale=0.5]
 \foreach \xpos in {0,0.5,1,2.3}
 {
   \draw[thick] (\xpos,0)--(\xpos,0.9);
   \draw[thick] (\xpos,1.1)--(\xpos,3);
 }
 \draw[thick] (-0.1,2) arc(90:270:0.5) -- (-0.1,1)--(2.4,1) arc(-90:90:0.5);
 \foreach \xpos in {0,0.5}
   \draw[thick] (\xpos+0.1,2.0)--(\xpos+0.4,2.0);
 \draw[thick] (1.1,2.0)--(2.2,2.0);
 \foreach \ypos in {0.5,2.5}
  \node at (1.7,\ypos) {$\cdots$};
\end{tikzpicture}
\end{equation}
where as usual the braids~\eqref{eq:g-diagr} are replaced by the combinations~\eqref{skein-rel}. It is not hard to check on the standard modules that the spectrum of $\C$ is non-degenerate in the sense of (R2). The subalgebras $\TL{N_1}$ and $\TL{N_2}$ have the corresponding central elements $\C^{(1)}$ and $\C^{(2)}$. Using diagrammatical manipulations one can actually compute eigenvalues of  $\C^{(1)}$ and $\C^{(2)}$ in $\StTL{j}[N]$ using the same type of arguments as below~\eqref{fuse-open-TL}, and this way to check the decomposition~\eqref{fuse-open-TL}.

To give an  example of (\ref{fuse-open-TL}), we have
\begin{eqnarray}
 \StTL{2}[8] \big|_{4+4} &=& \StTL{0}[4]\otimes\StTL{2}[4]~\oplus~\StTL{1}[4]\otimes\StTL{1}[4]~\oplus~\StTL{2}[4]\otimes\StTL{0}[4]\nonumber\\
 & \oplus & \StTL{1}[4]\otimes\StTL{2}[4]~\oplus~\StTL{2}[4]\otimes\StTL{1}[4]~\oplus~\StTL{2}[4]\otimes\StTL{2}[4] \,,
\label{fuse-open-TL-example}
\end{eqnarray}
where we use the short-hand notation $|_{N_1+N_2}$ for the restriction  $|_{\TL{N_1}\tensor\TL{N_2}}$.
In this formula, the first three direct summands correspond to zero connecting arcs, the next two summands to one connecting arc, and the last summand to two connecting arcs.
On the level of dimensions (\ref{fuse-open-TL-example}) corresponds to
$$
 20 = 2 \times 1 + 3 \times 3 + 1 \times 2 + 3 \times 1 + 1 \times 3 + 1 \times 1 \,,
$$
where we have used (\ref{dj}). This identity of course extends to the general case in~(\ref{fuse-open-TL}).

\section{Fusion for periodic TL in the generic case}\label{sec:generic}
In this section, we begin by giving several examples of the calculation of fusion rules for the standard modules $\StJTL{j}{z}$ for the affine TL algebra. These modules were introduced in Sec.~\ref{sec:St-mod}. We then set up an algebraic framework  leading to the general fusion rules in Sec.\ \ref{sec:gen-fus}, for generic parameters $\q$ and $z$. We finally discuss some of its properties: stability with $N$, associativity and braiding.

Contrary to the open case, the direct (i.e., using induction) diagrammatic calculation of the fusion is quite intricate. It is therefore limited to very small sizes of the system, as illustrated by the calculation on $1+1$ and $1+2$ sites in~\cite{GS}. The idea is then to use instead the more indirect way based on the Frobenius reciprocity~\eqref{eq:Frob-ATL}. For the open case this is quite straightforward, and was discussed above
in Sec.~\ref{sec:frob-fuse-open}. In the periodic case, this reciprocity amounts to the isomorphism
\begin{multline}
\label{eq:Frob-ATL-St}
\Hom_{\ATL{N}} \Bigl(\StJTL{j_1}{z_1}[N_1] \afus \StJTL{j_2}{z_2}[N_2], \StJTL{j}{z}[N]  \Bigr)  \\\cong
  \Hom_{\ATL{N_1}\tensor\ATL{N_2}} \Bigl( \StJTL{j_1}{z_1}[N_1] \tensor \StJTL{j_2}{z_2}[N_2], \StJTL{j}{z}[N] \big|_{N_1+N_2}\Bigr)\ ,
\end{multline}
where we now use the short-hand notation $|_{N_1+N_2}$ for the restriction  $|_{\ATL{N_1}\tensor\ATL{N_2}}$.
This isomorphism is best understood as our rule (R1), formulated in Sec.~\ref{sec:2rules}, which is valid for generic twists  $z_1$ and $z_2$. 
Our task is then to understand the decomposition of $\StJTL{j}{z}[N]$ with respect to the subalgebra $\ATL{N_1}\tensor\ATL{N_2}$ that was formulated in Sec.~\ref{sec:fusion-ATL}. In other words, we must find the multiplicities
\be\label{eq:ATL-St-decomp-gen}
\StJTL{j}{z}[N] \big|_{N_1+N_2} = \bigoplus_{(j_1,z_1),(j_2,z_2)} \mathsf{N}_{(j_1,z_1),(j_2,z_2)}^{(j,z)} \StJTL{j_1}{z_1}[N_1] \tensor \StJTL{j_2}{z_2}[N_2] \ .
\ee
 We will see below that for generic $z$ only the standard modules contribute to the decomposition. Some results for non-generic $z$ will be presented below. 
We also note that most of the multiplicities in~\eqref{eq:ATL-St-decomp-gen} are zero by dimension reasons: indeed, the left- and right-hand sides should have the same dimension. After determining the multiplicities we will finally be able to infer the fusion
\be\label{eq:afus-W-N}
 \StJTL{j_1}{z_1}[N_1] \afus \StJTL{j_2}{z_2}[N_2] = \bigoplus_{(j,z)} \mathsf{N}_{(j_1,z_1),(j_2,z_2)}^{(j,z)}\StJTL{j}{z}[N]
\ee
from the Frobenius reciprocity~\eqref{eq:Frob-ATL-St}. 
Meanwhile  it is easy to check that dimensions on both sides of~\eqref{eq:Frob-ATL-St} agree
by taking into account that 
\be
\dim \Hom_{\ATL{N_1}\tensor\ATL{N_2}} \bigl( \StJTL{k_1}{y_1}\tensor \StJTL{k_2}{y_2},\StJTL{j_1}{z_1}\tensor \StJTL{j_2}{z_2}\bigr) = \delta_{(k_1,y_1),(j_1,z_1)} \delta_{(k_2,y_2),(j_2,z_2)}\ 
\ee
for generic values of the twist parameters.

\subsection{Fusion on $1+1$ sites}\label{sec:ex}
We note that on $1$ site we have only the standard modules $\StJTL{\half}{z}[1]$ and the affine TL algebra is generated by $u$ only. In order to find fusion rules  of the standard modules $\StJTL{\half}{z_1}[1]$ and $\StJTL{\half}{z_2}[1]$, we follow the rule (R1) and begin with the decomposition for $\StJTL{0}{z}[2]$. Recall that the basis on $2$ sites is:
\be
\StJTL{0}{z}[2] = \Bigl\langle
a= 
\begin{tikzpicture}
 	\draw[thick, dotted] (-0.05,0.1) arc (0:10:0 and -3);
 	\draw[thick, dotted] (-0.05,0.1) -- (1.05,0.1);
 	\draw[thick, dotted] (1.05,0.1) arc (0:10:0 and -3);
	\draw[thick, dotted] (-0.05,-0.45) -- (1.05,-0.45);
	\draw[thick] (0.2,0.05) arc (-180:0:0.25 and 0.25);
	\end{tikzpicture}
\ ,\ 
 b=  \begin{tikzpicture}
 	\draw[thick, dotted] (-0.05,0.1) arc (0:10:0 and -3);
 	\draw[thick, dotted] (-0.05,0.1) -- (1.05,0.1);
 	\draw[thick, dotted] (1.05,0.1) arc (0:10:0 and -3);
	\draw[thick, dotted] (-0.05,-0.45) -- (1.05,-0.45);
	\draw[thick] (0,-0.2) arc (-90:0:0.25 and 0.25);
	\draw[thick] (1.0,-0.2) arc (-90:0:-0.25 and 0.25);
	\end{tikzpicture}
\Bigr\rangle
\ee
In this basis, we have then the following action of $u^{(1)}= u g_1^{-1}$
\be
u^{(1)} a = i \q^{\frac{3}{2}}b\ , \qquad 
u^{(1)} b = - i \q^{-\half} a + i \q^{\half}(z+z^{-1})b\ ,
\ee
while the action of $u^{(2)}=g_1 u$ reads
\be
u^{(2)} a = - i \q^{-\half}(z+z^{-1})a + i \q^{\half} b \ ,\qquad
u^{(2)} b = -i\q^{-\frac{3}{2}}a\ .
\ee
And we find that the spectrum of $u^{(1)}$ is $i \q^{\half}z^{\pm1}$ and  the spectrum of $u^{(2)}$ is $-i \q^{-\half}z^{\mp1}$, while $u^{(1)}u^{(2)}$ is  the identity.%
\footnote{This fact is enough to establish the decomposition, however one can also easily find the common eigenvectors: they are $v_{\pm}= a- \q z^{\pm1}b$.}
We thus infer the following decomposition
\be\label{eq:ex11-W0-decomp}
\StJTL{0}{z}[2]\big|_{1+1} =  \StJTL{\half}{i \q^{\half}z^{-1}}[1] \tensor \StJTL{\half}{-i \q^{-\half}z}[1] \; \oplus  \; 
 \StJTL{\half}{i \q^{\half}z}[1] \tensor \StJTL{\half}{-i \q^{-\half}z^{-1}}[1] \ .
\ee
We then similarly obtain the decomposition
\be\label{eq:ex11-W1-decomp}
\StJTL{1}{z}[2]\big|_{1+1} =  \StJTL{\half}{-i \q^{-\half}z}[1] \tensor \StJTL{\half}{i \q^{\half}z}[1] \ .
\ee

We can now apply the rule (R1) or, equivalently, \eqref{eq:afus-W-N} for this case. The two decompositions above then give the fusion rules: 
\begin{equation}\label{afus-ex1}
\StJTL{\half}{z_1}[1]\afus\StJTL{\half}{z_2}[1] =
\begin{cases}
\StJTL{1}{i\q^{\half} z_1}[2]\qquad & \qquad \text{when}\; z_2=-\q z_1,\\
\StJTL{0}{i\q^{\half}z_1^{-1}}[2]\qquad & \qquad \text{when}\; z_2=z_1^{-1},\\
0\qquad & \qquad \text{otherwise.}
\end{cases}
\end{equation}
We note that one term in~\eqref{eq:ex11-W0-decomp} corresponds to the resulting fusion $\StJTL{0}{z}$ while the other term to $\StJTL{0}{z^{-1}}$. However,  $\StJTL{0}{z}=\StJTL{0}{z^{-1}}$ by definition and this is why in the second line in~\eqref{afus-ex1} we have only one channel with $j=0$.
Of course, the result in~\eqref{afus-ex1} agrees with the direct calculation of induced modules in~\cite[Sec.\ 4.3.1]{GS}. However it is worth noticing that the calculation here is much simpler.
 In the fusion formula above, the result 0 on the right-hand side  means that there are no fusion channels, or, in other words, the momenta of left and right fusion factors are not compatible.
\medskip

We can of course do a similar analysis on higher number of sites but unfortunately the complexity grows very fast. Even on $1+2$ sites calculations are quite involved. It is clear that the direct study of  decomposition of standard modules for arbitrary $j$ onto standard modules over two ``small" algebras will be very complicated in general.
One would therefore need to find another analytical procedure that is more amenable to deal with the general case. Below we provide such an analysis based on spectral properties of some nice operators.

\subsection{Algebraic preliminaries}
\label{sec:alg-prelims}
We now set up the necessary machinery to compute the decompositions~\eqref{eq:ATL-St-decomp-gen} in  general cases for a given pair of quantum numbers $(j_1,z_1)$ and $(j_2,z_2)$.
We already noticed in Sec.~\ref{sec:2rules} that instead of (R1) we are going to use the more practical  rule (R2) which will be modified into studying the spectrum of a family of certain ``nice'' operators.

\subsubsection{The operators $\tau_j$}
\label{sec:tau-j}

An important role in the subsequent analysis will be played by the elements $\tau_j \in \ATL{N}$ defined by
\begin{equation}
 \tau_j = e_1 e_3 \cdots e_{N-1-2j-2} e_{N-1-2j} u \,,
\end{equation}
where we have supposed that $2j$ has the same parity as $N$ and satisfies $0 \le 2j \le N$. The corresponding diagram reads (here with $N=10$ and $j=2$):
\begin{equation}\label{tau-j-def}
 \begin{tikzpicture}
 \node  at (-0.7,-0.2) {$\tau_j =$};
 	\draw[thick, dotted] (-0.05,0.5) arc (0:10:0 and -7.5);
 	\draw[thick, dotted] (-0.05,0.55) -- (6.05,0.55);
 	\draw[thick, dotted] (6.05,0.5) arc (0:10:0 and -7.5);
	\draw[thick, dotted] (-0.05,-0.85) -- (6.05,-0.85);

	\draw[thick] (3.9,0.5) arc (0:10:40 and -7.6);
	\draw[thick] (4.5,0.5) arc (0:10:40 and -7.6);
	\draw[thick] (5.1,0.5) arc (0:10:40 and -7.6);
	\draw[thick] (5.7,0.5) arc (0:10:40 and -7.6);

	\draw[thick] (2.7,0.5) arc (-180:0:0.3 and 0.3);
	\draw[thick] (1.5,0.5) arc (-180:0:0.3 and 0.3);
	\draw[thick] (0.3,0.5) arc (-180:0:0.3 and 0.3);

	\draw[thick] (0.3,-0.8) arc (180:90:-0.3 and 0.3);
	\draw[thick] (5.7,-0.8) arc (180:90:0.3 and 0.3);
	\draw[thick] (0.9,-0.8) arc (180:0:0.3 and 0.3);
	\draw[thick] (2.1,-0.8) arc (180:0:0.3 and 0.3);
\end{tikzpicture}
\end{equation}
We can similarly define $\tau_j^{(i)}$ for subsystem $i=1,2$ by replacing $N$ by $N_i$ and $e_j$ by $e_j^{(i)}$.

To determine the spectrum of $\tau_j$, we first discuss its eigenvectors in the basis of link patterns.
A link pattern $v$ that enters the linear combination forming an eigenvector must necessarily have $N/2-j$ arcs in the same positions as the
arcs along the top rim of (\ref{tau-j-def}). This leaves $2j$ free points in $v$ upon which we can place any diagram of the standard module
$\StJTL{k}{z}[2j]$. But with respect to those free points, $\tau_j$ acts precisely as the translation operator $u$ for a system of size $2j$ points
only. It follows that the spectrum of $\tau_j$ in $\StJTL{k}{z}[N]$ coincides with the spectrum of $u$ in $\StJTL{k}{z}[2j]$:
\begin{equation}
 {\rm spec} \left(\tau_j,\StJTL{k}{z}[N]\right) = {\rm spec} \left(u,\StJTL{k}{z}[2j]\right) \,.
\end{equation}

This observation leads to some simple properties that we shall use extensively below:
\begin{enumerate}
 \item[(P1)] For $k > j$, $\tau_j$ is zero.
 \item[(P2)] For $j=0$, $\tau_j$ has a unique eigenvector of non-zero eigenvalue in $\StJTL{0}{z}[N]$, and this eigenvalue is $z+z^{-1}$.
 \item[(P3)] For $j > 0$ and $0 \le k \le j$, the eigenvalues of $\tau_j$ are of the form $z^{k/j} {\rm e}^{{\rm i}\pi \ell / j}$ with $\ell = 0,1,\ldots,2j-1$.
 (This follows from $u^{2j} = z^{2k}$.)
 \item[(P4)] For $k = j > 0$, we have a stronger statement: $\tau_j$ has a unique eigenvector of non-zero eigenvalue in $\StJTL{k}{z}[N]$, and this eigenvalue is $z$.
\end{enumerate}
The properties (P2) and (P4) are particularly useful, since applying them to $\tau_j^{(1)}\in\ATL{N_1}$ and $\tau_j^{(2)}\in\ATL{N_2}$ serves to fix the momenta, $z_1$ and $z_2$, appearing on the right-hand
side of the decomposition of $\StJTL{k}{z}[N]$.

Using the information about the spectrum of $\tau_j$
we can now reformulate the rule (R2) as follows:
\begin{itemize}
\item[(R3)]
If \textit{$\StJTL{k}{z}[N]$ appears in the fusion $\StJTL{j_1}{z_1}[N_1]\fus \StJTL{j_2}{z_2}[N_2]$ 
then  the following  is true:
\begin{enumerate}
\item  the spectrum of  $\tau_{j_1}^{(1)}$ on $\StJTL{k}{z}[N]$  contains
 $z_1$  for non-zero $j_1$ and $z_1+z_1^{-1}$ otherwise, and the multiplicity is divisible by $\hat d_{j_2}[N_2]$,
\item  the spectrum of  $\tau_{j_2}^{(2)}$ on $\StJTL{k}{z}[N]$  contains $z_2$  for non-zero $j_2$ and $z_2+z_2^{-1}$ otherwise, and the multiplicity is divisible by $\hat d_{j_1}[N_1]$,
\item  the spectrum of  $\tau_{j_1}^{(1)}\tau_{j_2}^{(2)}$ on $\StJTL{k}{z}[N]$  contains $z_1z_2$   for non-zero $j_1$ and $j_2$, or if otherwise $z_i$ should be replaced by $z_i+z_i^{-1}$ for the case $j_i=0$.
\end{enumerate}
}
\end{itemize}

Clearly this rule is weaker than the rule (R2): firstly, it is formulated for the standard modules only and secondly we provide only the necessary condition. The point is that analysis of the decomposition of   $\StJTL{k}{z}[N]$ using the spectral properties of $\tau_j^{(1)}$ and $\tau_j^{(2)}$ is rather involved.
Below we will use a combination of (R3) with more detailed analysis based on counting dimensions and taking into account the property (P3) formulated above.

\medskip

We now exhibit an analysis for the decomposition using the  $\tau_j^{(1)}$ and $\tau_j^{(2)}$ operators in an example on $4+4$ sites following the rule (R3).
This is intended to convince the reader that we can in fact find the decomposition of $\StJTL{k}{z}[N]$ for any
size $N$ such that the spectral analysis of the $\tau_j$ operators is feasible (using in practice a symbolic
algebra program such as {\sc Mathematica}).

\subsection{Example of the decomposition}
We begin with the practical considerations for general size systems.
 This will involve a few algebraic ingredients which are  studied systematically 
using exact symbolic, but computer-aided computations. Then we present  a specific example on
how the decomposition can be inferred for a rather large system with $N_1 + N_2 = 4 + 4$ sites. More exhaustive
results are deferred to Appendix~\ref{sec:appA}. These examples form a base that will lead us below to state a general conjecture, valid for
any $(N_1,N_2)$, which is supported by all the cases that we have worked out in details.

\subsubsection{Practical details}

We fix $N = N_1 + N_2$ and the $\ATL{N}$ module $\AStTL{j}{z}[N]$ appearing on the left-hand side of
the decomposition. We suppose that explicit representations of the first TL generator $e_1$ and the 
translation operator $u$ are given, both of dimension $\hat{d}_j[N]$, cf.\ (\ref{dim-dj}). The weight
of a contractible (resp.\ non-contractible) loop is parametrized as $m = \q + \q^{-1}$ (resp.\ $\widetilde{m} = z + z^{-1}$).

We first compute all the $e_j$, using the defining relations (\ref{TLpdef-u2}).
The braid operators $g_j$ and their inverses $g_j^{-1}$ are then given in terms of these by the
skein relations (\ref{skein-rel}).
The next step is to construct representations of the algebras $\ATL{N_1}$ and $\ATL{N_2}$ describing the two periodic subsystems. We denote
their generators as $e_j^{(1)}$ and $e_k^{(2)}$ respectively. First, the ordinary (open) TL generators are simply
defined by the embedding (\ref{TL-gens-embedding}).
The extra periodic generators, written graphically in (\ref{e0-1-def})--(\ref{e0-2-def}), are then obtained
from the algebraic expressions (\ref{e01-e02-alg}).
We shall also need the translation operators corresponding to the two subsystems, represented graphically in (\ref{u1-def})--(\ref{u2-def}), and defined algebraically by (\ref{trans-op-subsystems})
At this stage, one can obviously check on the given representation that all the defining relations are indeed satisfied.

\medskip

We have written a computer program in {\sc C++} that generates an explicit representation of $e_1$ and $u$ in $\StJTL{k}{z}[N]$.
These representations are written out as explicit matrices---with entries that depend on the variables $m$, $\widetilde{m}$ and $z$---in a file format which can 
subsequently be read into a symbolic algebra program such as {\sc Mathematica}. This step can be accomplished in a few minutes for sizes $N \le 20$, which is
already far larger than the sizes for which the remaining steps can be handled.

We next use {\sc Mathematica} to go through the algebraic steps (R3) in Section~\ref{sec:alg-prelims}. In particular we construct the operators $\tau_j^{(1)}$ and $\tau_j^{(2)}$
in terms of the variables $\q$ and $z$ (recall that $m = \q + \q^{-1}$ and $\widetilde{m} = z + z^{-1}$). These operators are then symbolically diagonalised. This step
is feasible in a reasonable time for sizes $N \le 12$ and all pairs $(N_1,N_2)$ such that $N_1 + N_2 = N$.

Using the properties of section~\ref{sec:tau-j} we can then infer the decomposition of $\StJTL{k}{z}[N]$. We now illustrate this procedure on a concrete example.

\subsubsection{Decomposition on $4+4$ sites}

Let us consider the case of $\StJTL{0}{z}[8]$ that we want to decompose on $N_1 + N_2 = 4 + 4$ sites. The dimension is $\hat{d}_0[8] = 70$.

We first find that
\begin{equation}
 {\rm spec} \left(\tau_0^{(1)},\StJTL{0}{z}[8]\right) = {\rm spec} \left(\tau_0^{(2)},\StJTL{0}{z}[8]\right) = \left\lbrace 0 \ (\times 64), z + z^{-1} \ (\times 6) \right\rbrace \,,
\end{equation}
where the multiplicities are shown between parentheses. Similarly, we find the spectrum of $\tau_0^{(1)}\tau_0^{(2)}$ which has only one non-zero eigenvalue $z + z^{-1}$.
This together with properties (P1)--(P2) implies that the decomposition contains a piece $\StJTL{0}{z_1}[4] \otimes \StJTL{0}{z_2}[4]$ with $z_1 = z^{\pm 1}$ and $z_2 = z^{\pm 1}$,
of dimension $6^2 = 36$, which matches the multiplicity $\hat{d}_0[4] = 6$ observed above.
It is not necessary to determine the relative signs in $z_1$ and $z_2$, since $\StJTL{0}{z}$ and
$\StJTL{0}{z^{-1}}$ are isomorphic. We have therefore
\begin{equation}
 \StJTL{0}{z}[8] = \StJTL{0}{z}[4] \otimes \StJTL{0}{z}[4] + \ldots \,,
 \label{W0z8_piece0}
\end{equation}
where the remaining pieces must have dimension $70 - 36 = 34$.

The spectra of the $\tau_1$ operators are:
\begin{eqnarray}
 {\rm spec} \left(\tau_1^{(1)},\StJTL{0}{z}[8]\right) = \left\lbrace 0 \ (\times 50), 1 \ (\times 6), -1 \ (\times 6), -\frac{\q}{z} \ (\times 4), - \q z \ (\times 4) \right \rbrace \,, \nonumber \\
 {\rm spec} \left(\tau_1^{(2)},\StJTL{0}{z}[8]\right) = \left\lbrace 0 \ (\times 50), 1 \ (\times 6), -1 \ (\times 6), -\frac{z}{\q} \ (\times 4), - \frac{1}{\q z} \ (\times 4) \right \rbrace \,.
\end{eqnarray}
By (P3), the eigenvalues $\pm 1$ come from the ${\cal W}_0$ piece that we have already determined in (\ref{W0z8_piece0}). The same piece is responsible for $6^2 - 12 = 24$ of
the zero eigenvalues. The other non-zero eigenvalues dictate, by (P4), another piece of the decomposition:
\begin{equation}
 \StJTL{1}{z_1}[4] \otimes \StJTL{1}{z_2}[4] \oplus \StJTL{1}{z_3}[4] \otimes \StJTL{1}{z_4}[4] \,,
 \label{W0z8_piece1}
\end{equation}
where $\{z_1,z_3\} = \{ -\frac{\q}{z},-\q z \}$ and $\{z_2,z_4\} = \{ -\frac{z}{\q},- \frac{1}{\q z}  \}$. Due to the uniqueness part of (P4) the piece is also responsible for
$2 \times 4 (4-1) = 24$ zero eigenvalues. There remains thus $50-24-24 = 2$ zero eigenvalues, and these must be attributed to tensorands ${\cal W}_k$ with $k > 1$, by (P1).
We shall come back to this below. Meanwhile, to correctly pair $\{z_1,z_3\}$ with $\{z_2,z_4\}$ we examine
\begin{equation}
 {\rm spec} \left(\tau_1^{(1)} \tau_1^{(2)},\StJTL{0}{z}[8]\right) =  \left\lbrace 0 \ (\times 64), 1 \ (\times 4), -1 \ (\times 2) \right\rbrace \,.
\end{equation}
It suffices here to notice that the spectrum is independent of $z$ (and of $\q$).
The product $+1$ must thus be associated with (\ref{W0z8_piece1}), whose final form
can therefore be fixed as
\begin{equation}
 \StJTL{1}{-\frac{\q}{z}}[4] \otimes \StJTL{1}{-\frac{z}{\q}}[4] \oplus \StJTL{1}{-\q z}[4] \otimes \StJTL{1}{-\frac{1}{\q z}}[4] \,.
 \label{W0z8_piece1final}
\end{equation}

The sum of the right-hand side of (\ref{W0z8_piece0}) and (\ref{W0z8_piece1final}) has dimension $6^2 + 2 \times 4^2 = 68$, so the last remaining piece of the decomposition
must take the form
\begin{equation}
 \StJTL{2}{z_5}[4] \otimes \StJTL{2}{z_6}[4] \oplus \StJTL{2}{z_7}[4] \otimes \StJTL{2}{z_8}[4] \,,
 \label{W0z8_piece2}
\end{equation}
to match the total dimension $\hat{d}_0[8] = 70$, seeing that $\hat{d}_2[4] = 1$. To find the corresponding momenta we examine $\tau_2^{(1)}$ and $\tau_2^{(2)}$. These
operators have rather complicated spectra, with no zero eigenvalues. While these can all be accounted for by the careful application of (P1)--(P4), it suffices here to notice
that they each have only two simple eigenvalues:
\begin{eqnarray}
 {\rm spec} \left(\tau_2^{(1)},\StJTL{0}{z}[8]\right) = \left\lbrace \frac{\q^2}{z} \ (\times 1), \q^2 z \ (\times 1), \ldots \right \rbrace \,, \nonumber \\
 {\rm spec} \left(\tau_2^{(2)},\StJTL{0}{z}[8]\right) = \left\lbrace \frac{z}{\q^2} \ (\times 1), \frac{1}{\q^2 z} \ (\times 1), \ldots \right \rbrace \,.
\end{eqnarray}
By (P4), it follows that in (\ref{W0z8_piece2}) we must have $\{z_5,z_7\} = \{ \frac{\q^2}{z}, \q^2 z \}$ and $\{z_6,z_8\} = \{ \frac{z}{\q^2},\frac{1}{\q^2 z}  \}$.
To obtain the correct pairing, we notice that the spectrum of $\tau_2^{(1)} \tau_2^{(2)}$ consists of 4th-roots of unity, with in particular no $z$-dependence.
This is sufficient to fix the final form of (\ref{W0z8_piece2}) as
\begin{equation}
 \StJTL{2}{\frac{\q^2}{z}}[4] \otimes \StJTL{2}{\frac{z}{\q^2}}[4] \oplus \StJTL{2}{\q^2 z}[4] \otimes \StJTL{2}{\frac{1}{\q^2 z}}[4] \,.
 \label{W0z8_piece2final}
\end{equation}

Summarizing the contents of (\ref{W0z8_piece0}), (\ref{W0z8_piece1final}) and (\ref{W0z8_piece2final}), we have established the decomposition
\begin{eqnarray}
 \StJTL{0}{z}[8] &=& \StJTL{0}{z}[4] \otimes \StJTL{0}{z}[4] \oplus \StJTL{1}{-\frac{\q}{z}}[4] \otimes \StJTL{1}{-\frac{z}{\q}}[4] \oplus \StJTL{1}{-\q z}[4] \otimes \StJTL{1}{-\frac{1}{\q z}}[4]  \nonumber \\
 &\oplus&  \StJTL{2}{\frac{\q^2}{z}}[4] \otimes \StJTL{2}{\frac{z}{\q^2}}[4] \oplus \StJTL{2}{\q^2 z}[4] \otimes \StJTL{2}{\frac{1}{\q^2 z}}[4] \,.
 \label{W0z8_final}
\end{eqnarray}
Here and in the sequel we omit the notation $|_{N_1+N_2}$ on the left-hand side whenever it can be inferred by inspecting the right-hand side.

\subsection{General results on the decomposition  of $\AStTL{j}{z}$}

Using these same ingredients, and some patience, it its possible to systematically determine the decomposition for any values of $N = N_1 + N_2$ for which the
{\sc Mathematica} computations are feasible. As an example, we present the complete results for $N=6$ and any values of $N_1$ and $N_2$ in Appendix~\ref{sec:appA}.

Studying carefully this body of results, we can conjecture the following final result, valid for any $N$:
\begin{equation}
 \AStTL{j}{z}[N_1+N_2] = \bigoplus_{j_1,j_2} \AStTL{j_1}{z_1}[N_1] \otimes \AStTL{j_2}{z_2}[N_2] 
\label{general_ATL_decomposition}
\end{equation}
with the following momenta:
\begin{itemize}
 \item For $j = j_1 + j_2$ and any values of $j_1,j_2 \ge 0$:
 \begin{equation}
  z_1 = (i \sqrt{\q})^{-2 j_2} z^{+1} \,, \qquad
  z_2 = (i \sqrt{\q})^{+2 j_1} z^{+1} \,;
  \label{j1+j2 case}
  \end{equation}
 \item For $j = j_1 - j_2$ and $j_1 \ge j_2 > 0$: 
 \begin{equation}
   \label{j1-j2 case}
  z_1 = (i \sqrt{\q})^{+2 j_2} z^{+1} \,, \qquad
  z_2 = (i \sqrt{\q})^{-2 j_1} z^{-1} \,;
  \end{equation}
 \item For $j = j_2 - j_1$ and $j_2 \ge j_1 > 0$: 
 \begin{equation}
  \label{j1-j2 case2}
  z_1 = (i \sqrt{\q})^{+2 j_2} z^{-1} \,, \qquad
  z_2 = (i \sqrt{\q})^{-2 j_1} z^{+1} \,;
  \end{equation}
\end{itemize}
Note that $N_1$ and $N_2$ do not appear explicitly  in these formulas---although of course the possible values of $j_1,j_2$ on the right hand side depend on them
(since we have the limitations $0 \le 2j_1 \le N_1$ and $0 \le 2j_2 \le N_2$).

We have checked this result exhaustively for $N \le 8$ using the algorithm with $\tau_j$ operators.
The formula above can be also tested analyticaly.
Firstly, we have verified that for any $N$ the total dimension of the right-hand side of 
(\ref{general_ATL_decomposition}) is equal to that of the left-hand side, as given by (\ref{dim-dj}).
Secondly, it agrees with the general result for one-dimensional representations we give below in~\eqref{eq:decompN-trivial}. 

Moreover, we can also {\em prove} the presence of some direct summands in the general decomposition formula. This is the case for the $j=j_1+j_2$ channel which we now consider in details, providing explicitly the eigenstates of $\tau_{j_1}^{(1)}$ and  $\tau_{j_2}^{(2)}$.

\subsubsection{Proof of the $j=j_1+j_2$ case}\label{sec:j1+j2 case}
We have been able to establish rigorously the presence of the direct summands corresponding to the $j=j_1+j_2$ channel~\eqref{j1+j2 case}. 
The idea of our proof is the following: we first note that in order to  detect a submodule $\StJTL{j_1}{z_1}[N_1] \otimes \StJTL{j_2}{z_2}[N_2]$ inside  $\StJTL{j}{z}[N]$ (for yet unknown $z_1$ and $z_2$) it is sufficient to provide common eigenvectors of $\tau_{j_1}^{(1)}$ and $\tau_{j_2}^{(2)}$ of non-zero eigenvalues; once they are found the eigenvalues are $z_1$ and $z_2$, correspondingly, in accordance with properties (P2) and (P4) established in Section~\ref{sec:tau-j}. For the case $j_1+j_2=j$, we were able to find such eigenvectors explicitly: these are the product (or rather the concatenation) of the corresponding eigenvectors in $\StJTL{j_1}{z_1}[N_1]$ and $\StJTL{j_2}{z_2}[N_2]$. An easy calculation then gives the eigenvalues $z_1$ and $z_2$ as in~\eqref{j1+j2 case}.

We illustrate such a calculation in several cases and  begin with the case $j=0$, and therefore set $j_1=j_2=0$. 
Let us denote  the state in $\StJTL{0}{z}[N]$ consisting of non-nested arcs by $v_0$:
\be
v_0\;=\;  \begin{tikzpicture}[scale=1/3, baseline = {(current bounding box.center)},rotate = 180]
	\draw[black, line width = 1pt] (0,0) .. controls (0,1) and (1,1) .. (1,0);
	\draw[black, line width = 1pt] (2,0) .. controls (2,1) and (3,1) .. (3,0);
	\node[anchor = north] at (4.5,0) {$\hdots $};
	\draw[black, line width = 1pt] (6,0) .. controls (6,1) and (7,1) .. (7,0);
	\draw[black, line width = 1pt] (8,0) .. controls (8,1) and (9,1) .. (9,0);
	\end{tikzpicture}\; .
\ee
This is the unique eigenvector, up to a scalar, with non-zero eigenvalue $z+z^{-1}$ of $\tau_0$ (if $z\ne e^{i\pi/2}$).
We then consider the action of $\tau_{0}^{(1)}$ and $\tau_{0}^{(2)}$ on $v_0$. 
Recall that 
\begin{equation}
 \tau_0^{(1)} = e_1 e_3 \cdots e_{N_1-1} u^{(1)} \,, \qquad
 \tau_0^{(2)} = e_{N_1+1} e_{N_1+3} \cdots e_{N-1} u^{(2)} 
\end{equation}
and $u^{(i)}$ are defined in~\eqref{trans-op-subsystems}, and diagrammatically in~\eqref{u1-def} and~\eqref{u2-def}.
Then a straightforward calculation in terms of diagrams (we skip it) shows that both the eigenvalues on $v_0$ equal to $z+z^{-1}$, and hence~\eqref{j1+j2 case} holds for $j=j_1=j_2=0$ for general $N_1$ and $N_2$. We only note that during the diagrammatical manipulations we  used the relations~\cite{GS}
\begin{equation}
g_i g_{i+1} e_i =  e_{i+1} e_i,\qquad g_i^{-1} g^{-1}_{i+1} e_i =  e_{i+1} e_i
\end{equation}
that tell us  that a  TL arc (``half'' of the diagram for $e_i$) can be pulled out under or above any string at the price of the factor~$1$.
Finally, it is also clear that we have identified the whole module $\StJTL{0}{z}[N_1] \otimes \StJTL{0}{z}[N_2]$ as it is generated from $v_0$ by the action of the subalgebra $\ATL{N_1}\tensor\ATL{N_2}$.

We now turn to the case of non-zero $j$ and when both $j_1$ and $j_2$ are non-zero too. We first recall   
that the eigenvector $v_{j}$ 
of $\tau_j$ acting on  $\StJTL{j}{z}[N]$ is a link state filled with non-nested arcs from left till the position $N-2j$ and the rest of sites are free, i.e., have $2j$ through-lines:
\be
v_j[N] \; = \;
	\begin{tikzpicture}[scale=1/3, baseline = {(current bounding box.center)},yscale=-1]
	\draw[black, line width = 1pt] (0,0) .. controls (0,1) and (1,1) .. (1,0);
	\node[anchor = north] at (2.5,0) {$\hdots$};
	\draw[black, line width = 1pt] (4,0) .. controls (4,1) and (5,1) .. (5,0);
	\draw[black, line width = 1pt] (6,0) .. controls (6,1) and (5,2) .. (5,3);
	\node[anchor = north] at (7.5,0) {$\hdots$};
	\node[anchor = south] at (6.5,3) {$\hdots$};
	\draw[black, line width = 1pt] (9,0) .. controls (9,1) and (8,2) .. (8,3);
\draw [black, line width = 1pt,decorate,decoration={brace,amplitude=5pt,mirror}] (4.8,3.3) -- (8.2,3.3) node[black,midway,yshift=-0.4cm] {\scriptsize $2j$};
	\end{tikzpicture} \ \ .
\ee
Its eigenvalue is non-zero and equals $z$; all the other (linearly independent) eigenvectors of $\tau_j$ come with~$0$ eigenvalue.
Now, in order to identify a submodule of the form $\StJTL{j_1}{z_1}[N_1] \otimes \StJTL{j_2}{z_2}[N_2]$ we consider the concatenation of $v_{j_1}[N_1]$ with $v_{j_2}[N_2]$  and calculate its image under the action of   $\tau_{j_1}^{(1)}$   and $\tau_{j_2}^{(2)}$. It turns out that this vector is a common eigenvector for both the operators with the corresponding eigenvalues $z_1$ and $z_2$ as in~\eqref{j1+j2 case}.
The crucial part of the calculation with  $\tau_{j_1}^{(1)}$  is to observe that within $\StJTL{j}{z}[N]$ we have (here the leftmost string goes over $2j_2$ through-lines)
\be
\begin{tikzpicture}[scale=1/3,baseline={(current bounding box.center)},yscale=-1]
	\foreach \r in {1,2,5,6}{
		\draw[black, line width = 1pt] (\r  ,0) .. controls (\r ,1) and (\r, 2) .. (\r,3);
	};
	\node[anchor = north] at (3.5,0) {$\hdots$};
	\node[anchor = south] at (3.5,3) {$\hdots$};
	\draw[white, line width = 3pt] (6.5,1) .. controls (6,1.5) and (0,2) .. (0,3);
	\draw[black, line width = 1pt] (6.5,1) .. controls (6,1.5) and (0,2) .. (0,3);
	\draw[black, line width = 1pt] (-0.5,1) .. controls (-0.2,0.8) and (-0.1,0.7) .. (0,0);
	\draw [black, line width = 1pt,decorate,decoration={brace,amplitude=5pt,mirror}] (0.7,3.3) -- (6.2,3.3) node[black,midway,yshift=-0.4cm] {\scriptsize $2j_2$};
\end{tikzpicture}
\; = \; \bigl(i\sqrt{\q}\bigr)^{-2j_2} z \;\times\;
\begin{tikzpicture}[scale=1/3,baseline={(current bounding box.center)},yscale=-1]
	\foreach \r in {0,1,2,5,6}{
		\draw[black, line width = 1pt] (\r  ,0) .. controls (\r ,1) and (\r, 2) .. (\r,3);
	};
		\node[anchor = north] at (3.5,0) {$\hdots$};
	\node[anchor = south] at (3.5,3) {$\hdots$};
		\draw [black, line width = 1pt,decorate,decoration={brace,amplitude=5pt,mirror}] (0.0,3.3) -- (6.2,3.3) node[black,midway,yshift=-0.4cm] {\scriptsize $2j_2+1$};
\end{tikzpicture}
\ee
which is due to the fact that all but one term close  through-lines and so are zero. The value in this equation is indeed $z_1$ as in~\eqref{j1+j2 case}. A similar reasoning works for $z_2$.
The only remaining case that we have to analyse is when $j$ is non-zero but one of $j_1$ and $j_2$ is zero. The calculation is similar to the above one, so we skip it, and the result again agrees with~\eqref{j1+j2 case}. This finishes our proof. 

By counting dimensions, it is clear that the contribution of the case $j=j_1+j_2$ as in~\eqref{j1+j2 case} does not cover the whole space  $\StJTL{j}{z}[N]$. There should be contributions with $j_1 + j_2 > j$ but $|j_1-j_2|=j$. The idea here is that the ``extra" $(j_1 + j_2 - j)$ through-lines are paired/joined into arcs in a non-trivial way in order to provide an embedding of $\StJTL{j_1}{z_1} \otimes \StJTL{j_2}{z_2}$ to $\StJTL{j}{z}$  in this case. We made the analysis in few low-dimension cases and the rules for such pairings are quite involved, and a general pattern for them remains unclear to us.

\subsection{General affine TL fusion}
\label{sec:gen-fus}
From the above formula~\eqref{general_ATL_decomposition}, we have  the decompositions in the form~\eqref{eq:ATL-St-decomp-gen}. Using the Frobenius reciprocity reformulated by~\eqref{eq:afus-W-N}, we can now read off the only non-zero results for our affine TL fusion:
\begin{subequations}
 \label{genfusii}
 \begin{align}
 {\rm Plus\ channel: } &
 \begin{cases}
 \ \ \ \, \AStTL{j_1}{z_1} \afus \AStTL{j_2}{(i\sqrt{\q})^{2(j_1+j_2)}z_1} = \AStTL{j_1+j_2}{(i\sqrt{\q})^{2j_2}z_1}\,, & \ \ \, \; j_1,j_2 \ge 0 \label{genfusii:plus}
 \end{cases} \\[0.5mm]
 {\rm Minus \ channels:} &
 \begin{cases}
 \AStTL{j_1}{z_1} \afus \AStTL{j_2}{(i\sqrt{\q})^{2(-j_1+j_2)}z^{-1}_1} = \AStTL{j_1-j_2}{(i\sqrt{\q})^{-2j_2}z_1}\,, & j_1 \ge j_2 > 0 \\
 \AStTL{j_1}{z_1} \afus \AStTL{j_2}{(i\sqrt{\q})^{2(-j_1+j_2)}z^{-1}_1} = \AStTL{-j_1+j_2}{(i\sqrt{\q})^{2j_2}z^{-1}_1}\,, & j_2 \ge j_1 > 0 \label{genfusii:minus}
 \end{cases}
\end{align}
\end{subequations}

We note here that the fusion results do not depend on the values of $N_1,N_2$ (provided they allow for the  values of $j$ under consideration) so we suppress mention of these in the following. Of course, the result here agrees with the previous calculations in Section~\ref{sec:ex} and in~\cite{GS}.

One may also choose to rewrite the fusion rules \eqref{genfusii} so that the left-hand side reads simply $\AStTL{j_1}{z_1} \afus \AStTL{j_2}{z_2}$ in
all cases, while the right-hand side involves appropriate conditions on $z_1$ and $z_2$. Some care should however be taken in such a rewriting.
Indeed, when either one or both of $j_1$, $j_2$ are zero---notice that this can happen only in the plus channel \eqref{genfusii:plus}---we have to take
into account the identification between $\AStTL{0}{z}$ and $\AStTL{0}{z^{-1}}$ when reading off the fusion rules from the branching rules
\eqref{j1+j2 case}--\eqref{j1-j2 case2}.

Consider as an example the branching rule \eqref{eq:A6} 
 according to which $\AStTL{1}{z}[N]$ contains not only $\AStTL{0}{-z/\q}[N_1] \otimes \AStTL{1}{z}[N_2]$,
but also, by the identification, $\AStTL{0}{-\q/z}[N_1] \otimes \AStTL{1}{z}[N_2]$. Therefore the fusion $\AStTL{0}{z_1} \afus \AStTL{1}{z_2}$ produces, in the plus channel
\eqref{genfusii:plus}, not only $\AStTL{1}{-\q z_1}$ if $z_2/z_1 = -\q$, but also $\AStTL{1}{-\q/z_1}$ if $z_1 z_2 = -\q$.
So the fusion formula is symmetric under $z_1 \to z_1^{-1}$ as it should be taking into account the identification $\AStTL{0}{z_1} = \AStTL{0}{z_1^{-1}}$.

In the general case, we thus arrive at the following alternative form of the fusion: 
\begin{eqnarray}
 \AStTL{j_1}{z_1} \afus \AStTL{j_2}{z_2} &=&
 \delta_{z_2/z_1, (i \sqrt{\q})^{2(j_1+j_2)}} \AStTL{j_1+j_2}{(i \sqrt{\q})^{2 j_2} z_1}  \nonumber \\
 &+& \left( \delta_{j_1,0} + \delta_{j_2,0} - \delta_{j_1,0} \delta_{j_2,0} \right) \delta_{z_1 z_2,(i \sqrt{\q})^{2(-j_1+j_2)}} \AStTL{j_1+j_2}{(i \sqrt{\q})^{2 j_2} z_1^{-1+2\delta_{j_2,0}}} \nonumber \\
 &+& \delta_{j_1 \ge j_2 > 0} \delta_{z_1 z_2,(i\sqrt{\q})^{2(-j_1+j_2)}} \AStTL{j_1-j_2}{(i\sqrt{\q})^{-2j_2}z_1} \nonumber \\
 &+& \delta_{j_2 \ge j_1 > 0} \delta_{z_1 z_2,(i\sqrt{\q})^{2(-j_1+j_2)}} \AStTL{-j_1+j_2}{(i\sqrt{\q})^{2j_2}z^{-1}_1} \,.
\end{eqnarray}
 In this formula, the first two lines account for the plus channel (taking into account the identification $\AStTL{0}{z} = \AStTL{0}{z^{-1}}$ as
just discussed), while the last two lines provide the minus channels (that do not present any subtleties, since they only occur for $j_1,j_2 > 0$).
We shall however not use this rewriting of the fusion rules further below, since the original form \eqref{genfusii:plus} is much easier to write and manipulate.

\subsection{Properties}
We study here certain properties of the affine TL fusion formulated above, such as stability with~$N$, associativity and braiding.

\subsubsection{Stability with $N$}\label{sec:stability}
As we already mentioned above, the fusion rules~\eqref{genfusii} for the standard modules do not depend on number of sites they are defined.
This conclusion agrees with the stability property proven in~\cite[Prop.\,6.7]{GS},  using a functorial construction involving the so-called  globalization functors---they  respect the fusion while changing the value of $N$ but not $j$. This property says that the fusion rules, or numbers $ \mathsf{N}_{(j_1,z_1),(j_2,z_2)}^{(j,z)}$ in~\eqref{eq:afus-W-N}, for a fixed pair  $(j_1,z_1)$ and $(j_2,z_2)$ do not depend on the choice of $(N_1,N_2)$. Using this and our symbolic calculation for the limited amount of cases for $N\leq 8$ we have actually proven many fusion rules which are only limited by certain $j_1,j_2$ and not by the size of the system. 

Using the stability property, we can also establish  the ``plus" channel in~\eqref{genfusii:plus}.
Indeed, let us first extend the result of the decomposition~\eqref{eq:ex11-W1-decomp} to the  case of one-dimensional standard modules on any number of sites. After a similar calculation we get
\begin{equation}\label{eq:decompN-trivial}
 \AStTL{\frac{N}{2}}{z}[N]\big|_{N_1+N_2}=\AStTL{{N_1\over2}}{(i\sqrt{\q})^{-N_2}z}[N_1]\tensor \AStTL{{N_2\over 2}}{(i\sqrt{\q})^{N_1}z}[N_2] \ .
\end{equation}
This comes from the action of $u^{(1)}$ on the single state in $\AStTL{N/2}{z}$, where one should keep only the term not containing $e_j$'s, and similarly for $u^{(2)}$.
This formula allows us to deduce  the fusion rule of an extreme case with $j_1=N_1/2$ and $j_2=N_2/2$:
\be
\AStTL{{N_1\over2}}{(i\sqrt{\q})^{-N_2}z}[N_1]\afus \AStTL{{N_2\over 2}}{(i\sqrt{\q})^{N_1}z}[N_2] =  \AStTL{\frac{N_1 + N_2}{2}}{z}[N] \ .
\ee
 Applying to this formula our stability property---i.e., replacing $j$-indexes $N_1\over2$ and $N_2\over2$ by $j_1$ and $j_2$, respectively, while choosing the number of sites in square brackets $N'_1>N_1$ and $N'_2>N_2$---we  get the general formula for the plus channel:
\be
\AStTL{j_1}{(i\sqrt{\q})^{-2j_2}z}[N'_1]\afus \AStTL{j_2}{(i\sqrt{\q})^{2j_1}z}[N'_2] =  \AStTL{j_1+j_2}{z}[N'_1+N'_2] \ .
\ee
This clearly agrees with~\eqref{genfusii:plus} and also confirms the discussion in Section~\ref{sec:j1+j2 case}.

\subsubsection{Associativity}
That the affine TL fusion is associative was proven in~\cite{GS}, and here we provide  some checks that the general formulas~\eqref{genfusii} we proposed do indeed agree with the associativity of $\afus$. 
We start by discussing fusion in the plus channel \eqref{genfusii:plus}.
We consider the  fusion of three standard modules:
\begin{equation}\label{eq:fus-triple-gen}
\AStTL{j_1}{z_1}\afus \AStTL{j_2}{z_2}\afus \AStTL{j_3}{z_3}\ .
\end{equation}
The result of this fusion is non trivial in the plus channel  only if 
\begin{eqnarray}\label{eq:fus-assoc-plus-gen}
{z_1\over z_2}=(i\sqrt{\q})^{-2j_1-2j_2}\ ,\nonumber\\
{z_2\over z_3}=(i\sqrt{\q})^{-2j_2-2j_3}\ .
\end{eqnarray}
If~\eqref{eq:fus-assoc-plus-gen} holds, the fusion of the second and third modules in~\eqref{eq:fus-triple-gen} gives $\AStTL{j_2+j_3}{(i\sqrt{\q})^{2j_3}z_2}$. The fusion of this module is possible with $\AStTL{j_1}{z_1}$ if and only if ${z_1/((i\sqrt{\q})^{2j_3}z_2})=(i\sqrt{\q})^{-2j_1-2(j_2+j_3)}$. But this is the same as the compatibility for fusion of the first and second modules. This means that, if the fusion of first and second modules is non-zero in the plus channel but not so for the fusion of second and third modules, or the fusion of second and third modules is non-zero in the plus channel  but not so for the fusion of first and second modules, the result will be zero, which is  compatible with associativity.

From this analysis, we will thus consider 
\begin{equation}\label{eq:3W}
\AStTL{j_1}{z}\afus \AStTL{j_2}{(i\sqrt{\q})^{2j_1+2j_2}z}\afus \AStTL{j_3}{(i\sqrt{\q})^{2j_1+4j_2+2j_3} z} \ ,
\end{equation}
where the labels are such that all fusions give non trivial results. If we first fuse the two modules on the left we get
\begin{equation}
\AStTL{j_1+j_2}{(i\sqrt{\q})^{2j_2}z}\afus \AStTL{j_3}{(i\sqrt{\q})^{2j_1+4j_2+2j_3} z}=\AStTL{j_1+j_2+j_3}{(i\sqrt{\q})^{2j_2+2j_3}z}\ .
\end{equation}
Meanwhile, if  in~\eqref{eq:3W} we do fusion of the last two modules first then we get
\begin{equation}
\AStTL{j_1}{z}\afus \AStTL{j_2+j_3}{(i\sqrt{\q})^{2j_1+2j_2+2j_3} z}=\AStTL{j_1+j_2+j_3}{(i\sqrt{\q})^{2j_2+2j_3}z}
\end{equation}
in agreement with the previous formula.

\medskip

Consider as another example the fusion that involves a combination of the minus and plus channels~\eqref{genfusii} 
%
\begin{equation}
\AStTL{j_1}{z}\afus \AStTL{j_2}{(i\sqrt{\q})^{2j_2-2j_1}z^{-1}}\afus \AStTL{j_3}{(i\sqrt{\q})^{2j_1-4j_2+2j_3}z}\label{plusminus} \ ,
\end{equation}
where the labels are again chosen to satisfy coherence conditions.
Fusion of the first two modules in the minus channel~\eqref{genfusii:minus}  produces 
\begin{equation}
\AStTL{j_1-j_2}{(i\sqrt{\q})^{-2j_2}z}\afus \AStTL{j_3}{(i\sqrt{\q})^{2j_1-4j_2+2j_3}z}\label{resi}\ .
\end{equation}
In order for this fusion to be possible in the minus channel again, we need to have 
\begin{equation}
(i\sqrt{\q})^{-2j_2}z\times (i\sqrt{\q})^{2j_1-4j_2+2j_3}z=(i\sqrt{\q})^{2j_3-2j_1+2j_2}
\end{equation}
forcing $z=(i\sqrt{q})^{-2j_1+4j_2}$, so (\ref{resi}) becomes
\begin{equation}
\AStTL{j_1-j_2}{(i\sqrt{\q})^{-2j_1+2j_2}}\afus \AStTL{j_3}{(i\sqrt{\q})^{2j_3}}=\AStTL{j_1-j_2-j_3}{(i\sqrt{\q})^{-2j_1+2j_2-2j_3}}
\label{fusfinalj1-j2-j3}
\end{equation}
since the  product of the two $z$ variables involved allows for fusion in the minus channel. Now, we go back to (\ref{plusminus}) but consider instead the fusion of the second and third modules. This fusion will be possible in the plus channel if and only if 
\begin{equation}
{(i\sqrt{\q})^{2j_2-2j_1}z^{-1}\over (i\sqrt{\q})^{2j_1-4j_2+2j_3}z}=(i\sqrt{\q})^{-2j_2-2j_3}
\end{equation}
which forces $z=(i\sqrt{\q})^{-2j_1+4j_2}$ again. We then get 
\begin{equation}
\AStTL{j_1}{(i\sqrt{\q})^{-2j_1+4j_2}}\afus \AStTL{j_2+j_3}{(i\sqrt{\q})^{2j_3-2j_2}}=\AStTL{j_1-j_2-j_3}{(i\sqrt{\q})^{-2j_1+2j_2-2j_3}}
\end{equation}
which is consistent with fusion in the minus channel and gives the same result as (\ref{fusfinalj1-j2-j3}).
Meanwhile, the fusion of the second and third modules in (\ref{plusminus}) is generically possible in the minus channel, giving rise to 
\begin{equation}
\AStTL{j_1}{z}\afus \AStTL{j_2-j_3}{(i\sqrt{\q})^{2j_2-2j_1-2j_3}z^{-1}}=\AStTL{j_1-j_2+j_3}{(i\sqrt{\q})^{-2j_2+2j_3}z}
\end{equation}
where for the equality we used  fusion in the minus channel again. On the other hand, we can go back to (\ref{resi}) and observe that fusion in the plus channel is always possible, giving 
\begin{equation}
\AStTL{j_1-j_2}{(i\sqrt{\q})^{-2j_2}z}\afus \AStTL{j_3}{(i\sqrt{\q})^{2j_1-4j_2+2j_3}z}=\AStTL{j_1-j_2+j_3}{(i\sqrt{\q})^{-2j_2+2j_3}z}
\end{equation}
again confirming associativity. 

\subsubsection{Braiding}\label{sec:br}
Equations (\ref{genfusii}) show that our fusion rules are 
non commutative: this is ultimately due to the asymmetry between the first (or left) and the second (or right) subalgebras in our construction, since for the generators of the first subalgebra one goes around the system by passing {\sl over} the $N_2$ strands on the right, while for the second subalgebra, one passes {\sl below} the $N_1$ strands of the first chain. It is of course possible to define fusion the other way, interchanging above and below. This corresponds to using $g_i$ (resp.\ $g_i^{-1}$) instead of $g_i^{-1}$ (resp.\ $g_i$) in the maps of translation operators, giving rise to (recall~\eqref{utilde1-def} and the discussion above it)
\begin{equation}
\tilde{u}^{(1)}=ug_{N-1}\ldots g_{N_1},~~~\tilde{u}^{(2)}=g_{N_1}^{-1}\ldots g_1^{-1}u
\end{equation}
We will denote the corresponding fusion by $\afusm$, its properties with respect to $\afus$ were studied in~\cite{GS}.
The final result for corresponding fusion rules is obtained simply by replacing $i\sqrt{\q}$ by $(i \sqrt{\q})^{-1}$, as we do same replacing when changing braids to inverse braids, leading to  
\begin{subequations}
 \label{genmodfusii}
 \begin{align}
 {\rm Plus\ channel: } &
 \begin{cases}
 \, \AStTL{j_1}{z_1} \afusm \AStTL{j_2}{(i\sqrt{\q})^{-2(j_1+j_2)}z_1} = \AStTL{j_1+j_2}{(i\sqrt{\q})^{-2j_2}z_1}\,, & \ \ \, \; j_1,j_2 \ge 0 \label{genmodfusii:plus}
 \end{cases} \\[0.5mm]
 {\rm Minus \ channels:} &
 \begin{cases}
 \AStTL{j_1}{z_1} \afusm \AStTL{j_2}{(i\sqrt{\q})^{2(j_1-j_2)}z^{-1}_1} = \AStTL{j_1-j_2}{(i\sqrt{\q})^{2j_2}z_1}\,, & j_1 \ge j_2 > 0 \\
 \AStTL{j_1}{z_1} \afusm \AStTL{j_2}{(i\sqrt{\q})^{2(j_1-j_2)}z^{-1}_1} = \AStTL{-j_1+j_2}{(i\sqrt{\q})^{-2j_2}z^{-1}_1}\,, & j_2 \ge j_1 > 0 \label{genmodfusii:minus}
 \end{cases}
\end{align}
\end{subequations}
It is straightforward to check this is exactly the same as the foregoing fusion, with the left and right spaces interchanged
\begin{equation}\label{eq:fusm-iso}
\AStTL{j_1}{z_1} \afusm \AStTL{j_2}{z_2}\cong\AStTL{j_2}{z_2}\afus \AStTL{j_1}{z_1}\ ,
\end{equation}
in agreement with~\cite{GS}. 
It is also easy to see that the  two fusions can be related by using the braid operator which passes strings from the left over those from the right:
\begin{equation}\label{sigma-def}
 \begin{tikzpicture}
 \node
   at (-1.5,-4) {\mbox{} $g_{N_1,N_2}\;\;\equiv$ \mbox{}\qquad};
\braid[braid colour=black,strands=5,braid width=\brw,braid start={(-0.3,-0.88)}]  {\g_2^{-1}\g_1^{-1}\g_3^{-1}\g_2^{-1}\g_4^{-1}\g_3^{-1}}

\node[font=\large]  at (4.0,-4) {\mbox{}\qquad \(=\)};
\draw[thick] (5.5,-2.5) -- (6.2,-3.9);
\draw[thick] (5.75,-2.5) -- (6.3,-3.6);

\draw[thick](6.65,-4.6)--(7.1,-5.5);
\draw[thick] (6.80,-4.4) -- (7.35,-5.5);

\draw[thick] (5.5,-5.5) -- (7,-2.5);
	\draw[thick] (5.75,-5.5) -- (7.25,-2.5);
         \draw[thick] (6,-5.5) -- (7.5,-2.5);
\end{tikzpicture}
\end{equation}
or defined formally by
\begin{equation}
g_{N_1, N_2} = \left(g_{N_2}^{-1}\ldots g_2^{-1}g_1^{-1}\right) \left(g_{N_2+1}^{-1}\ldots g_2^{-1}\right)\ldots \left(g_{N_2+N_1-1}^{-1}\ldots g_{N_1}^{-1}\right)\ .
\end{equation}
It relates the two embeddings:
\begin{align}
g_{N_1,N_2} \cdot u^{(1)}[N_1] \cdot g_{N_1,N_2}^{-1}&=\tilde{u}^{(2)}[N_2]\nonumber\\
g_{N_1,N_2} \cdot u^{(2)}[N_2] \cdot g_{N_1,N_2}^{-1}&=\tilde{u}^{(1)}[N_2]\ 
\end{align}
and thus provides an explicit isomorphism (via the action) in~\eqref{eq:fusm-iso}, see more details in~\cite{GS}.

\section{Fusion in the (partially) degenerate case}\label{sec:partdeg}
We now discuss the case where, while $\q$ remains generic, $z$ is not any longer.
It is well known~\cite{GL}  that when  the standard modules $\AStTL{j}{z}$ are not irreducible they are indecomposable and have a unique irreducible quotient. In more detail, the standard module   $\AStTL{j'}{z'}$ has a non-zero homomorphism to another standard module  $\AStTL{j}{z}$ 
\begin{equation}\label{cell-emb}
\StJTL{j'}{z'}\hookrightarrow\StJTL{j}{z}
\end{equation}
if and only if  $j'-j=k$ for a non-negative integer $k$ and the pairs $(j',z')$ and $(j,z)$ satisfy
\begin{equation}\label{eq:emb-cond}
z'  = z(-\q)^{-\epsilon k} \qquad \text{and}\qquad z^2 = (-\q)^{\epsilon 2j'},\qquad \text{for} \quad  \epsilon=\pm1\ .
\end{equation}
Note that we then have $(z')^2 = (-\q)^{2\epsilon j} $. When $\q$ is not a root of unity, there is at most one solution to (\ref{cell-emb}). When there is one, the module $\StJTL{j}{z}$ has a unique proper irreducible submodule isomorphic to $\StJTL{j'}{z'}$. One can then obtain a simple module by taking the quotient
\begin{equation}
\bAStTL{j}{z}\equiv\StJTL{j}{z}/\StJTL{j'}{z'}\label{Wbar-def}
\end{equation}
We now discuss how these quotients behave under the fusion we have defined earlier.

Taking now our general result (\ref{general_ATL_decomposition}) for given, one can form the quotient (\ref{Wbar-def}) 
 on the left-hand side.
We use then the fact that restriction of the quotient (to a subalgebra) is the quotient of restrictions.%
\footnote{Mathematically this says that the restriction functor is exact.}
 We then need just to combine properly the terms on the right-hand side using (\ref{general_ATL_decomposition}) and the rules in~\eqref{eq:emb-cond} in order to kill the corresponding terms and thus get the quotient in the end.
In the following we suppose $N$ even, whence $j$ is integer. Systematically computing the right-hand sides for different values of $N = N_1 + N_2$, e.g.\
as large as $N = 24$, and reorganising the direct sums, we have established, for instance, that 
\begin{eqnarray}
 \overline{\cal W}_{j,i^{2j}\q^{j+1}} &=& \!\! \bigoplus_{0 \le \ell \le j} \overline{\cal W}_{\ell,i^{2 \ell}\q^{\ell+1}} \otimes
 \left( {\cal W}_{j-\ell,i^{2(j+\ell)}\q^{j+\ell+1}} \ominus {\cal W}_{j+\ell+1,i^{2(j-\ell-1)}\q^{j-\ell}} \right) \label{conjecture1_first} \\
 &+& \bigoplus_{\ell > j} \overline{\cal W}_{\ell,i^{2\ell}\q^{\ell+1}} \otimes
 \left( {\cal W}_{\ell-j,i^{-2(j+\ell)} \q^{-(j+\ell+1)}} \ominus {\cal W}_{j+\ell+1,i^{2(j-\ell-1)} \q^{j-\ell}} \right) \,. \nonumber
\end{eqnarray}
for the $k=1$ case. 
We have here used the notation $\ominus$ instead of the usual quotient sign; the reason is that
these expressions do not yet have the correct form in the sense of the irreducible quotient modules.
For the more general situation with $k \ge 1$ integer we similarly find that
\begin{eqnarray}
 \overline{\cal W}_{j,i^{2j}\q^{j+k}} &=& \!\! \bigoplus_{0 \le \ell \le j} \overline{\cal W}_{\ell,i^{2 \ell}\q^{\ell+k}} \otimes
 \left( {\cal W}_{j-\ell,i^{2(j+\ell)}\q^{j+\ell+k}} \ominus {\cal W}_{j+\ell+k,i^{2(j-\ell-k)}\q^{j-\ell}} \right) \label{conjecture_k_first} \\
 &+& \bigoplus_{\ell > j} \overline{\cal W}_{\ell,i^{2\ell}\q^{\ell+k}} \otimes
 \left( {\cal W}_{\ell-j,i^{-2(j+\ell)} \q^{-(j+\ell+k)}} \ominus {\cal W}_{j+\ell+k,i^{2(j-\ell-k)} \q^{j-\ell}} \right) \nonumber \\
 &+& \!\!\!\!\! \bigoplus_{0 < \ell < k/2}'
 \left( {\cal W}_{\ell,i^{2 \ell} \q^{\ell-k}} \ominus {\cal W}_{k - \ell,i^{2(k-\ell)} \q^{-\ell}} \right) \otimes
 \left( {\cal W}_{j+\ell,i^{2(j+\ell)} \q^{k+j-\ell}} \ominus {\cal W}_{k+j-\ell,i^{2(k+j-\ell)} \q^{j+\ell}} \right) \,. \nonumber
\end{eqnarray}
In the third sum, the mark signifies that $\ell$ is understood to be
integer (resp.\ half an odd integer) if $N_1$ is even (resp.\ odd). In particular, for $k=1$ this sum is empty, and so
(\ref{conjecture_k_first}) correctly reduces to (\ref{conjecture1_first}). The sum is also empty for $k=2$, provided that $N_1$ is even.

 Meanwhile,
we have also checked that the left- and right-hand sides of (\ref{conjecture1_first})--(\ref{conjecture_k_first}) have the same dimensions
for any values of $N$. 
This makes us confident that they are generally true (and in particular for both parities of $N$).

For $j$ and $\ell$ integer we have
\begin{eqnarray*}
 {\cal W}_{j-\ell,i^{2(j+\ell)} \q^{j+\ell+k}} &=& {\cal W}_{j-\ell,i^{2(j-\ell)} \q^{j-\ell+2\ell+k}} \,, \\
 {\cal W}_{j+\ell+k,i^{2(j-\ell-k)} \q^{j-\ell}} &=& {\cal W}_{j-\ell+2\ell+k,i^{2(j+\ell+k)} \q^{j-\ell}} \,.
\end{eqnarray*}
Taking differences, and applying (\ref{Wbar-def}) with $j \to j-\ell$ and $k \to 2 \ell + k$, we see that the difference $\ominus$ in the first
line of (\ref{conjecture_k_first}) can be identified as 
$\bAStTL{j-\ell}{i^{2(j-\ell)} q^{j+\ell+k}}$. Similar manipulations on the remaining terms
lead to the general result of the decomposition in the degenerate case:
\begin{eqnarray}
 \bAStTL{j}{i^{2j}\q^{j+k}} &=& \!\! \bigoplus_{0 \le \ell \le j} \bAStTL{\ell}{ i^{2 \ell}\q^{\ell+k}}
 \otimes \bAStTL{j-\ell}{i^{2(j-\ell)}\q^{j+\ell+k}} 
  \label{conjecture_k} \\
 &+& \bigoplus_{\ell > j} \bAStTL{\ell}{i^{2\ell}\q^{\ell+k}}
\otimes\bAStTL{\ell-j}{i^{2(\ell-j)} \q^{-(j+\ell+k)}} \nonumber \\
 &+& \!\!\!\!\! \bigoplus_{0 < \ell < k/2}'
 \bAStTL{\ell}{ i^{2 \ell} \q^{\ell-k}} \otimes \bAStTL{j+\ell}{  i^{2(j+\ell)} \q^{k+j-\ell}}
 \,. \nonumber
\end{eqnarray}

Using the Frobenius reciprocity this corresponds to three fusion channels extending results in \eqref{j1+j2 case}--\eqref{j1-j2 case2} to the degenerate case
\begin{equation}
\bAStTL{j_1}{i^{2j_1}\q^{j_1+k}}\afus \bAStTL{j_2}{i^{2j_2}\q^{j_2+2j_1+k}}=\bAStTL{j_1+j_2}{i^{2(j_1+j_2)}\q^{j_1+j_2+k}}\label{fusion1}
\end{equation}
and 
\begin{equation}
\bAStTL{j_1}{i^{2j_1}\q^{j_1+k}}\afus\bAStTL{j_2}{i^{2j_2}\q^{j_2-k-2 j_1}} =\bAStTL{j_1-j_2}{i^{2(j_1-j_2)}\q^{j_1-j_2+k}}, \quad \mbox{for } j_1 \geq j_2 \label{fusion2}
\end{equation}
with, in addition, 
\begin{equation}
\bAStTL{j_1}{i^{2j_1}\q^{j_1-k}}\afus \bAStTL{j_2}{i^{2j_2}\q^{j_2+k-2j_1}}=\bAStTL{j_2-j_1}{i^{2(j_2-j_1)}\q^{j_2-j_1+k}},\quad \mbox{for } j_2\geq j_1\ .\label{fusion3}
\end{equation}
Similar formulas can be obtained for all the other solutions of the degeneracy equations~\eqref{eq:emb-cond}. 
However, we should mention that we are not at generic points and modules $\bAStTL{j}{z}$, for $z$ as in~\eqref{eq:emb-cond},  admit indecomposable but reducible projective covers and tilting modules. In principle, one would have to analyse branching rules for these modules too and see if $\bAStTL{j_1}{z_1}\tensor \bAStTL{j_2}{z_2}$ summands appear there too  or not (if they appear, then such modules should provide extra channels in the fusion). We nevertheless believe that such summands do not appear there and we thus did not miss anything.

\section{The conformal limit}\label{sec:CFTlimit}
As is well known, many models of physical interest---in particular, those providing lattice regularizations of Logarithmic Conformal Field Theories (LCFT)---are based on the Temperley Lieb-algebra, see e.g.~\cite{ReadSaleur07-2,PRZ,GV,GJRSV}. More precisely, the 
 transfer matrices and hamiltonians of these models are  particular elements of  the algebra $\TL{N}(m)$ with open boundary conditions and of the  $\ATL{N}(m)$ algebra with periodic (or ``twisted'') boundary conditions. In this case, a detailed analysis 
 going back more than 25 years \cite{AGR,PS} has established a correspondence between modules of the lattice algebras and modules of the corresponding Virasoro  (resp.\ product of two Virasoro) algebras in the continuum limit for the open (resp.\ periodic) cases. This correspondence was used, together with a lattice definition of fusion in the open case, to infer fusion in boundary LCFTs  in \cite{ReadSaleur07-1}, see also~\cite{GV} for more direct connection to logarithmic OPEs. Obviously, our definition of fusion in the periodic case should similarly give results for {\sl fusion of non-chiral fields in LCFTs}.
 
To start, we set $m=\q+\q^{-1}$, and restrict to the case $\q$ a complex number of unit modulus $\q={\rm e}^{i\pi\over x+1}$, with $x \in \mathbb{R}_+$. We also restrict to Hamiltonians of the form $H=-\sum_i e_i$, for which it is known that a theory with central charge
 \begin{equation}
 c=1-{6\over x(x+1)}
 \end{equation}
is obtained in the continuum limit. The scaling limit of modules can be inferred from the knowledge of the generating functions on systems of $N$ sites,
 \begin{equation}
  \hbox{Tr }{\rm e}^{-\beta_R(H-N\varepsilon_0)} e^{-i\beta_IP}\;\xrightarrow{\, N\to\infty\,}\; \hbox{Tr}\, q^{L_0-c/24}\bar{q}^{\bar{L}_0-c/24} \,, \label{charform}
 \end{equation}
 where $H$ is the lattice hamiltonian  (normalized such that the velocity of sound is unity), and $P$ is the lattice momentum, while $\varepsilon_0$ 
 is the ground state energy per site in the thermodynamic limit. We also set $q(\bar{q})=\exp\left[-{2\pi\over N}(\beta_R\pm i\beta_I)\right]$ with $\beta_{R,I} \in \mathbb{R}$ and $\beta_R>0$, and $N$ is the length of the chain.%
 \footnote{
 One should of course not confuse the modular parameters $q$ and $\bar{q}$ with
 the quantum-group related parameter~$\q$.} 

\subsection{The generic case}

We start by considering  the case where $\q$ and $z$ are both generic. We know \cite{AGR,PS}
 the trace in (\ref{charform}) when evaluated on the module $\AStTL{j}{z}$: 
\begin{equation}
F_{j,z}={(q\bar{q})^{-c/24}\over  P(q)P(\bar{q})}\sum_{n=-\infty}^\infty q^{{[xj+(x+1)(n+\phi)]^2-1\over 4x(x+1)}}
\bar{q}^{{[xj-(x+1)(n+\phi)]^2-1\over 4x(x+1)}}\label{generalcontlim}
\end{equation}
for $\q={\rm e}^{\frac{i \pi}{x+1}}$ and $z={\rm e}^{i\pi\phi}$ and here we also used $P(q) =  \prod_{n=1}^{\infty} (1 - q^n)$. Note moreover that the exponent in this sum can be written formally using the Kac formula 
\begin{equation}
h_{r,s}={[(x+1)r-xs]^2-1\over4x(x+1)}\ ,
\end{equation}
so the sum in (\ref{generalcontlim}) is over $q^{h_{n+\phi,-j}} \bar{q}^{h_{n+\phi,j}}$. We recognize in this trace the Virasoro  characters of the  Verma modules $\Verma{r,s}$ corresponding to the weight $h_{r,s}$:
\begin{equation}
\hbox{Tr}_{~\Verma{r,s}}~q^{L_0-c/24}={q^{h_{rs}-c/24}\over P(q)}\ .
\end{equation}
This, together with arguments based on the lattice discretization of the
Virasoro generators \cite{KooSaleur,GRS3, GRSV1,GJRSV}, indicates that the Virasoro algebra content of the continuum limit of the module $\AStTL{j}{z}$ is%
\begin{equation}
\AStTL{j}{z}\mapsto \bigoplus_{n\in \mathbb{Z}} \Verma{n+\phi,-j}\tensore \bVerma{n+\phi,j}
\end{equation}
where we used the notation $\tensore$ to denote the outer product of left and right  (chiral and antichiral) Virasoro representations.
We note that in this formula $\phi$ is manifestly $1$-periodic,  while the change $\phi\mapsto \phi\pm1$ leads to the change $z\mapsto -z$. However, the Virasoro algebra action depends only on  $z^2$, as can be seen by inspecting the Koo--Saleur formula~\cite{KooSaleur,GJRSV}. Indeed,  the Virasoro modes $L_n$ and $\bar{L}_n$  are expressed in terms of $e_j$'s only, and the action of the latter depends on $z^2$ and not on the sign of $z$.

We now discuss the continuum limit of the ATL fusion results. 
Consider first the plus channel, which is the case where the parameters satisfy the constraints (\ref{genfusii:plus}). 
Writing $z=e^{i\pi \phi}$ and $z_i=e^{i\pi \phi_i}$ we can reformulate~(\ref{genfusii:plus}) as conditions on non-zero fusion: $\phi_1=-j_2-{j_2\over x+1}+\phi$ and $\phi_2=j_1+{j_1\over x+1}+\phi$. This allows us some flexibility to rewrite the continuum limit of the modules $ \AStTL{j}{z}$ involved. For instance we have 
\begin{eqnarray}
F_{j_1,z_1} &=& { (q\bar{q})^{-c/24}\over P(q)P(\bar{q})}\sum_{n=-\infty}^\infty q^{{[xj_1+(x+1)(n+\phi-j_2-{j_2\over x+1})]^2-1\over 4x(x+1)}}
\bar{q}^{{[xj-(x+1)(n+\phi-j_2-{j_2\over x+1})]^2-1\over 4x(x+1)}}\nonumber\\
&=& { (q\bar{q})^{-c/24}\over P(q)P(\bar{q})}\sum_{n=-\infty}^\infty q^{{[xj_1+(x+1)(n+\phi-2j_2+{xj_2\over x+1})]^2-1\over 4x(x+1)}}
\bar{q}^{{[xj_1-(x+1)(n+\phi-2j_2+{xj_2\over x+1})]^2-1\over 4x(x+1)}}\nonumber\\
&=& {(q\bar{q})^{-c/24}\over  P(q)P(\bar{q})}\sum_{p=-\infty}^\infty q^{{[x(j_1+j_2)+(x+1)(p+\phi)]^2-1\over 4x(x+1)}}
\bar{q}^{{[x(j_1-j_2)-(x+1)(p+\phi)]^2-1\over 4x(x+1)}}
\end{eqnarray}
with $p=n-2j_2$, and thus, with the conditions (\ref{genfusii:plus}), we can rewrite
\begin{equation}
 \AStTL{j_1}{z_1}\mapsto \bigoplus_{n\in \mathbb{Z}} \Verma{n+\phi_1,-j_1}\tensore \bVerma{n+\phi_1,j_1}=
  \bigoplus_{p\in \mathbb{Z}} \Verma{p+\phi_,-j_1-j_2}\tensore \bVerma{p+\phi,j_1-j_2}
  \end{equation}
 A similar trick leads to 
\begin{equation}
 \AStTL{j_2}{z_2}\mapsto \bigoplus_{n\in \mathbb{Z}} \Verma{n+\phi_2,-j_2}\tensore \bVerma{n+\phi_2,j_2}=
  \bigoplus_{m\in \mathbb{Z}} \Verma{m+\phi_,j_1-j_2}\tensore \bVerma{m+\phi,j_1+j_2}
  \end{equation}
where now $m=n+2j_1$. This allows us to interpret the continuum limit of the fusion 
 in  the plus channel (\ref{genfusii:plus}) as
\begin{eqnarray}
\left(\bigoplus_p\Verma{p+\phi,-j_1-j_2}\tensore \bVerma{p+\phi,j_1-j_2}\right)\afusV\left(\bigoplus_m\Verma{m+\phi,j_1-j_2}\tensore
\bVerma{m+\phi,j_1+j_2}\right)\nonumber\\
=\bigoplus_r\Verma{r+\phi,-j_1-j_2}\tensore\bVerma{r+\phi,j_1+j_2}\label{nicefus}.
\end{eqnarray}

Observe that in (\ref{nicefus}) the result on the right-hand side is obtained by ``glueing'' the antichiral component from the first factor with the chiral one from the second one. 
This generalizes to the two other fusion channels. For instance, 
in the second case in  (\ref{genfusii})  we find
\begin{eqnarray}
\left(\bigoplus_n \Verma{n+\phi,-j_1+j_2}\tensore\bVerma{n+\phi,j_1+j_2}\right)\afusV\left(
\bigoplus_m \Verma{m+\phi,j_1+j_2}\tensore\bVerma{m+\phi,j_1-j_2}\right)\nonumber\\
=\bigoplus_r \Verma{r+\phi,-j_1+j_2}\tensore\bVerma{r+\phi,j_1-j_2}.
\end{eqnarray}

\subsection{The (partially) degenerate case}

Things  look somewhat nicer in the partially degenerate case where, while $\q$ (and hence $x$) remains generic, the momentum parameter $z$ takes degenerate values. In this case, the exponents in~\eqref{generalcontlim} are all in the extended Kac table. Moreover, the trace  (\ref{charform}) over the modules $\bAStTL{j}{i^{2j}q^{j+k}}$ can be reexpressed in terms of  characters of irreducible Virasoro modules, which are Kac modules. One finds indeed 
\begin{equation}
F_{j,i^{2j}q^{j+k}}^{(0)}=F_{j,i^{2j}q^{j+k}}-F_{j+k,i^{2(+k)j}q^j}=\sum_{r=1}^\infty K_{rk}\overline{K}_{r,k+2j}\label{Flimit}
\end{equation}
where the $K_{rs}$ functions are characters of the (Kac) irreducible Virasoro modules $\IrrV{r,s}$:
\begin{equation}
K_{r,s} = 
\hbox{Tr}_{~\IrrV{r,s}}~q^{L_0-c/24}={q^{h_{rs}}-q^{h_{r,-s}}\over q^{c/24}P(q)}\ .
\end{equation}
Similarly, we have 
\begin{equation}
F_{j,i^{2j}q^{-(j+k)}}^{(0)}=F_{j,i^{2j}q^{-(j+k)}}-F_{j+k,i^{2(j+k)}q^{-j}}=\sum_{r=1}^\infty K_{r,k+2j}\overline{K}_{r,k} \,,
\end{equation}
and we note that the chiral and antichiral components have been exchanged, as compared with (\ref{Flimit}). The conformal equivalent of (\ref{fusion1})  then reads
\begin{equation}
\left(\bigoplus_{r=1}^\infty \IrrV{r,k}\tensore \IrrVb{r,k+2j_1}\right)\afusV\left(
\bigoplus_{s=1}^\infty \IrrV{s,k+2j_1}\tensore\IrrVb{s,k+2(j_1+j_2)}\right)= \bigoplus_{t=1}^\infty \IrrV{t,1}\tensore\IrrVb{t,k+2(j_1+j_2)}\ .
\label{cfusion1}
\end{equation}
For (\ref{fusion2}) we have 
\begin{equation}
\left(\bigoplus_{r=1}^\infty \IrrV{r,k}\tensore\IrrVb{r,k+2j_1}\right)\afusV\left(\bigoplus_{s=1}^\infty \IrrV{s,k+2j_1}\tensore\IrrVb{s,k+2(j_1-j_2)}\right)= \bigoplus_{t=1}^\infty \IrrV{r,k}\tensore\IrrVb{r,k+2(j_1-j_2)}\ .
\label{cfusion2}
\end{equation}
Finally, for (\ref{fusion3})
\begin{equation}
\left(\bigoplus_{r=1}^\infty \IrrV{r,k}\tensore\IrrVb{r,k-2j_1}\right)\afusV\left(\bigoplus_{s=1}^\infty \IrrV{r,k-2j_1}\tensore\IrrVb{s,k+2(j_2-j_1)}\right)= \bigoplus_{t=1}^\infty \IrrV{r,k}\tensore\IrrVb{r,k+2(j_2-j_1)}\ .
\label{cfusion3}
\end{equation}
In all these cases, we have 
\begin{equation}
\left(\bigoplus_{r=1}^\infty \IrrV{r,a}\tensore\IrrVb{r,b}\right)\afusV\left(\bigoplus_{s=1}^\infty \IrrV{r,b}\tensore\IrrVb{r,c}\right)=\bigoplus_{t=1}^\infty \IrrV{t,a}\tensore\IrrVb{t,c}\label{mystpatt}\ .
\end{equation}

\subsection{Interpretation}

While the results we have obtained so far involve only infinite sums of modules, the basic principle of glueing between the antichiral sector of the first component in the tensor product and the chiral sector of the second component can obviously be performed on each term of the sums individually. We thus believe that the continuum limit interpretation of our results is 
\begin{equation}
(\Phi_{a}\tensore\bar{\Phi}_b)\afusV(\Phi_c\tensore\bar{\Phi}_d)=\delta_{bc} (\Phi_{a}\tensore\bar{\Phi}_d)\label{fusinter}
\end{equation}
where $\Phi,\bar{\Phi}$ denote primary fields of the CFT labelled by $a,b,c,d$. 
Equation (\ref{fusinter})  can be interpreted as the (balanced) tensor product over the Virasoro algebra: the non-chiral field $\Phi_{a}\tensore\bar{\Phi}_b$ generates a bimodule over (one copy of) Virasoro algebra, 
where the left Virasoro action is for the chiral part and the right action corresponds to the anti-chiral algebra, and both actions commute. 
Then taking the tensor product of the two bimodules over the algebra Vir is by definition (for the corresponding fields in them)
\begin{equation}
(\Phi_{a}\tensore\bar{\Phi}_b) \otimes_{\rm{Vir}} (\Phi_b\tensore\bar{\Phi}_c) \,,
\end{equation}
which is again a bimodule that corresponds to
\begin{equation}
\Phi_{a}\tensore\bar{\Phi}_c \,,
\end{equation}
at least for $\Phi_b$ corresponding to a simple module, and any $\Phi_a$ and $\Phi_c$.

The fusion is  non-commutative since, by exchanging the two terms in (\ref{fusinter}), we have
\begin{equation}
(\Phi_{c}\tensore\bar{\Phi}_d)\afusV(\Phi_a\tensore\bar{\Phi}_b)=\delta_{da} (\Phi_{c}\tensore\bar{\Phi}_b)\label{fusinteri}
\end{equation}
which is clearly not the same as (\ref{fusinter}). Meanwhile the fusion is associative:
\begin{eqnarray}
(\Phi_{a}\tensore\bar{\Phi}_b) \afusV \left((\Phi_b\tensore\bar{\Phi}_c)\afusV(\Phi_c\tensore\bar{\Phi}_d)\right)=
(\Phi_{a}\tensore\bar{\Phi}_b) \afusV (\Phi_b\tensore\bar{\Phi}_d)=(\Phi_{a}\tensore\bar{\Phi}_d)\nonumber\\
\left((\Phi_{a}\tensore\bar{\Phi}_b) \afusV (\Phi_b\tensore\bar{\Phi}_c)\right)\afusV(\Phi_c\tensore\bar{\Phi}_d)=
(\Phi_{a}\tensore\bar{\Phi}_c) \afusV (\Phi_c\tensore\bar{\Phi}_d)=(\Phi_{a}\tensore\bar{\Phi}_d)
\end{eqnarray}
Note that the interpretation in (\ref{fusinter}) explains a posteriori why the result of so many fusions is zero: a precise matching of the sectors is necessary for the glueing to be possible. 

\subsection{Comments}
There is nothing special about glueing the right movers from the first field in the fusion with the left mover from the second field. Indeed, recall the other fusion $\afusm$ introduced in Section~\ref{sec:br}: it is easy to see that, if we use $\afusm$ instead of $\afus$, the fusion rules (\ref{genmodfusii}) lead, for instance  to
\begin{eqnarray}
\left(\bigoplus_n \Verma{n+\phi,j_1-j_2}\tensore
\bVerma{n+\phi,-j_1-j_2}\right)\afusV^-\left(\bigoplus_p \Verma{p+\phi,j_1+j_2}\tensore\bVerma{p+\phi,j_1-j_2}\right)\nonumber\\
=\left(\bigoplus_r \Verma{r+\phi,j_1+j_2}\tensore\bVerma{r+\phi,-j_1-j_2}\right) \,.
\end{eqnarray}
 in the continuum limit. Here the left movers from the first field are now glued to the right movers of the second one.
This of course agrees with the braiding in~\eqref{eq:fusm-iso} relating both the fusions.

\section{Conclusion}\label{sec:Conclusion}

While we believe that our definition of fusion and the corresponding results for finite-dimensional ATL modules are interesting in their own right, it is clear that we have not obtained what one would like to call a ``lattice version'' of bulk fusion. What we found instead in the continuum limit is simply that the antichiral part of one component is glued to the chiral part of the other component, without any ``fusion'' in
the usual physical sense that would correspond to bulk operator product expansions. Extending this interpretation to the degenerate case leads to seemingly unphysical results. In the case of the Ising model for instance ($\q=e^{i\pi/4}$), we would get 
  that the fusion $\afus$ of simple modules corresponds, in the continuum limit, to 
\begin{eqnarray}
(I \tensore \bar{I}) \afusV (I \tensore \bar{I}) = (I \tensore \bar{I})\nonumber\\
(\epsilon\tensore \bar{\epsilon})\afusV(\epsilon\tensore \bar{\epsilon})=(\epsilon\tensore \bar{\epsilon})\nonumber\\
(\sigma\tensore \bar{\sigma})\afusV(\sigma\tensore \bar{\sigma})=(\sigma\tensore \bar{\sigma})\label{Isingfus}
\end{eqnarray}
where $I,\epsilon,\sigma$ are the three primary fields of the Ising ($c={1\over 2}$) CFT and antiholomorphic components are denoted by the usual bar.%
\footnote{This result can be recovered by direct calculations using Majorana fermions.}
Clearly, relations  (\ref{Isingfus}) are very different from the physical fusion rules \cite{DfMS}, which involve, in particular,
\begin{equation}
 (\sigma\tensore \bar{\sigma}) \times (\sigma\tensore \bar{\sigma}) = (I \tensore \bar{I})\oplus(\epsilon\tensore \bar{\epsilon}) \,,
\end{equation}
where $\times$ now means the ordinary bulk CFT fusion.

On the other hand, our way of defining fusion, based on embedding two ``small" algebras into a big one, seems essentially unique. We provide an analysis of the uniqueness issue in Appendix~\ref{sec:appB}.
We do not know if there is another way to define fusion that would have a more satisfactory continuum limit, for instance, using an approach that does not rely on an embedding.

We observe that the fusion results for finite-dimensional standard ATL modules suggest a possible fusion for the infinite-dimensional modules (i.e., those 
where configurations with through-lines wrapping around the cylinder several times are not reduced to those with the lines not wrapping~\cite{MartinSaleur}). Setting formally
\begin{equation}
\widehat{W}_j\sim \int dz ~\AStTL{j}{z}
\end{equation}
we immediately get from (\ref{genfusii}) 
\be
\widehat{W}_{j_1} \afus \widehat{W}_{j_2} =  \widehat{W}_{j_1+j_2} \oplus  \widehat{W}_{|j_1-j_2|}
\ee
which is very reminiscent of the fusion one would have in a Coulomb gas or in Liouville theory~\cite{CGLiouville}. 
The study of infinite-dimensional modules of ATL and the possible associated lattice models appears like one of the most interesting areas for future study.

An obvious application of our results concerns fusion for the blob algebra modules combining with idea of ``braid translation" from \cite{BGJSR}.

\medskip

{\bf Acknowledgments.} \hspace{3pt} We acknowledge interesting discussions with   Jonathan  Bellet\^ete, Vaughan Jones, Yvan St Aubin, and Zhenghan Wang.
This work was supported by the Agence Nationale de la Recherche (grant ANR-10-BLAN-0414), the Institut
Universitaire de France, the European Research Council (advanced grant NuQFT). We are also grateful to MATRIX Institute of Melbourne University and to the organizers of  the conference \textit{Integrability in low-dimensional Quantum Systems}  in Creswick in 2017 where a part of this work was done.
The work of AMG was supported by CNRS. AMG is also grateful to the Mathematics Department of Hamburg University for kind hospitality.

\appendix

\section{Decompositions of $\StJTL{j}{z}[N]$ for $N=6$}
\label{sec:appA}

As a compendium, we present here the complete results for the decompositions of $\StJTL{j}{z}[N]$ on $N=6$ sites.
All of these formula agree, of course, with the general conjecture (\ref{general_ATL_decomposition}).

\subsection{Results on $6=1+5$ sites}\label{app:A1}
\begin{eqnarray}
 \StJTL{0}{z}[6] &=& \StJTL{\frac12}{i \sqrt{q} z}[1] \otimes \StJTL{\frac12}{\frac{1}{i \sqrt{q} z}}[5] \oplus
   \StJTL{\frac12}{\frac{i \sqrt{q}}{z}}[1] \otimes \StJTL{\frac12}{\frac{z}{i \sqrt{q}}}[5] \\ 
 \StJTL{1}{z}[6] &=& \StJTL{\frac12}{\frac{z}{i \sqrt{q}}}[1] \otimes \StJTL{\frac12}{i \sqrt{q} z}[5] \oplus
   \StJTL{\frac12}{\frac{(i \sqrt{q})^3}{z}}[1] \otimes \StJTL{\frac32}{\frac{z}{i \sqrt{q}}}[5] \\  
 \StJTL{2}{z}[6] &=& \StJTL{\frac12}{\frac{z}{(i \sqrt{q})^3}}[1] \otimes \StJTL{\frac32}{i \sqrt{q} z}[5] \oplus
   \StJTL{\frac12}{\frac{(i \sqrt{q})^5}{z}}[1] \otimes \StJTL{\frac52}{\frac{z}{i \sqrt{q}}}[5] \\ 
 \StJTL{3}{z}[6] &=& \StJTL{\frac12}{\frac{z}{(i \sqrt{q})^5}}[1] \otimes \StJTL{\frac52}{i \sqrt{q} z}[5] 
\end{eqnarray}

\subsection{Results on $6=2+4$ sites}\label{app:A2}
\begin{eqnarray}
 \StJTL{0}{z}[6] &=& \StJTL{0}{z}[2] \otimes \StJTL{0}{z}[4] \oplus \nonumber \\
   & & \StJTL{1}{(i \sqrt{q})^2 z}[2] \otimes \StJTL{1}{\frac{1}{(i \sqrt{q})^2 z}}[4] \oplus \StJTL{1}{\frac{(i \sqrt{q})^2}{z}}[2] \otimes \StJTL{1}{\frac{z}{(i \sqrt{q})^2}}[4] \\ 
 \StJTL{1}{z}[6] &=& \StJTL{0}{\frac{z}{(i \sqrt{q})^2}}[2] \otimes \StJTL{1}{z}[4] \oplus \nonumber \\
   & & \StJTL{1}{z}[2] \otimes \StJTL{0}{\frac{1}{(i \sqrt{q})^2 z}}[4] \oplus \StJTL{1}{\frac{(i \sqrt{q})^4}{z}}[2] \otimes \StJTL{2}{\frac{z}{(i \sqrt{q})^2}}[4] \label{eq:A6} \\ 
 \StJTL{2}{z}[6] &=& \StJTL{0}{\frac{z}{(i \sqrt{q})^4}}[2] \otimes \StJTL{2}{z}[4] \oplus \StJTL{1}{\frac{z}{(i \sqrt{q})^2}}[2] \otimes \StJTL{1}{(i \sqrt{q})^2 z}[4] \\ 
 \StJTL{3}{z}[6] &=& \StJTL{1}{\frac{z}{(i \sqrt{q})^4}}[2] \otimes \StJTL{2}{(i \sqrt{q})^2 z}[4] 
\end{eqnarray}

\subsection{Results on $6=3+3$ sites}\label{app:A3}
\begin{eqnarray}
 \StJTL{0}{z}[6] &=& \StJTL{\frac12}{i \sqrt{q} z}[3] \otimes \StJTL{\frac12}{\frac{1}{i \sqrt{q} z}}[3] \oplus \StJTL{\frac12}{\frac{i \sqrt{q}}{z}}[3] \otimes \StJTL{\frac12}{\frac{z}{i \sqrt{q}}}[3] \oplus \nonumber \\
  & & \StJTL{\frac32}{(i \sqrt{q})^3 z}[3] \otimes \StJTL{\frac32}{\frac{1}{(i \sqrt{q})^3 z}}[3] \oplus \StJTL{\frac32}{\frac{(i \sqrt{q})^3}{z}}[3] \otimes \StJTL{\frac32}{\frac{z}{(i \sqrt{q})^3}}[3] \\ 
 \StJTL{1}{z}[6] &=& \StJTL{\frac12}{\frac{z}{i \sqrt{q}}}[3] \otimes \StJTL{\frac12}{i \sqrt{q} z}[3] \oplus \nonumber \\
  & & \StJTL{\frac32}{i \sqrt{q} z}[3] \otimes \StJTL{\frac12}{\frac{1}{(i \sqrt{q})^3 z}}[3] \oplus \StJTL{\frac12}{\frac{(i \sqrt{q})^3}{z}}[3] \otimes \StJTL{\frac32}{\frac{z}{i \sqrt{q}}}[3] \\ 
 \StJTL{2}{z}[6] &=& \StJTL{\frac32}{\frac{z}{i \sqrt{q}}}[3] \otimes \StJTL{\frac12}{(i \sqrt{q})^3 z}[3] \oplus \StJTL{\frac12}{\frac{z}{(i \sqrt{q})^3}}[3] \otimes \StJTL{\frac32}{i \sqrt{q} z}[3] \\ 
 \StJTL{3}{z}[6] &=& \StJTL{\frac32}{\frac{z}{(i \sqrt{q})^3}}[3] \otimes \StJTL{\frac32}{(i \sqrt{q})^3 z}[3]
\end{eqnarray}

\subsection{Results on $6=4+2$ sites}\label{app:A4}
\begin{eqnarray}
 \StJTL{0}{z}[6] &=& \StJTL{0}{z}[4] \otimes \StJTL{0}{z}[2] \oplus \nonumber \\
  & & \StJTL{1}{(i \sqrt{q})^2 z}[4] \otimes \StJTL{1}{\frac{1}{(i \sqrt{q})^2 z}}[2] \oplus \StJTL{1}{\frac{(i \sqrt{q})^2}{z}}[4] \otimes \StJTL{1}{\frac{z}{(i \sqrt{q})^2}}[2] \\ 
 \StJTL{1}{z}[6] &=& \StJTL{0}{\frac{z}{(i \sqrt{q})^2}}[4] \otimes \StJTL{1}{z}[2] \oplus \nonumber \\
  & & \StJTL{1}{z}[4] \otimes \StJTL{0}{(i \sqrt{q})^2 z}[2] \oplus \StJTL{2}{(i \sqrt{q})^2 z}[4] \otimes \StJTL{1}{\frac{1}{(i \sqrt{q})^4 z}}[2] \\ 
 \StJTL{2}{z}[6] &=& \StJTL{1}{\frac{z}{(i \sqrt{q})^2}}[4] \otimes \StJTL{1}{(i \sqrt{q})^2 z}[2] \oplus
    \StJTL{2}{z}[4] \otimes \StJTL{0}{\frac{1}{(i \sqrt{q})^4 z}}[2] \\ 
 \StJTL{3}{z}[6] &=& \StJTL{2}{\frac{z}{(i \sqrt{q})^2}}[4] \otimes \StJTL{1}{(i \sqrt{q})^4 z}[2]
\end{eqnarray}

\subsection{Results on $6=5+1$ sites}\label{app:A5}
\begin{eqnarray}
 \StJTL{0}{z}[6] &=&  \StJTL{\frac12}{i \sqrt{q} z}[5] \otimes \StJTL{\frac12}{\frac{1}{i \sqrt{q} z}}[1] \oplus
    \StJTL{\frac12}{\frac{i \sqrt{q}}{z}}[5] \otimes \StJTL{\frac12}{\frac{z}{i \sqrt{q}}}[1] \\
 \StJTL{1}{z}[6] &=& \StJTL{\frac12}{\frac{z}{i \sqrt{q}}}[5] \otimes \StJTL{\frac12}{i \sqrt{q} z}[1] \oplus
    \StJTL{\frac32}{i \sqrt{q} z}[5] \otimes \StJTL{\frac12}{\frac{1}{(i \sqrt{q})^3 z}}[1] \\
 \StJTL{2}{z}[6] &=& \StJTL{\frac32}{\frac{z}{i \sqrt{q}}}[5] \otimes \StJTL{\frac12}{(i \sqrt{q})^3 z}[1] \oplus
    \StJTL{\frac52}{i \sqrt{q} z}[5] \otimes \StJTL{\frac12}{\frac{1}{(i \sqrt{q})^5 z}}[1] \\
 \StJTL{3}{z}[6] &=& \StJTL{\frac52}{\frac{z}{i \sqrt{q}}}[5] \otimes \StJTL{\frac12}{(i \sqrt{q})^5 z}[1]
\end{eqnarray}

We note that the results in Sections~\ref{app:A2} and~\ref{app:A4}, as well as in Sections~\ref{app:A1} and~\ref{app:A5}, are the same after disregarding the dependence on $N_1$ and $N_2$, in line with the general result given in Section~\ref{sec:stability}.

\section{Search for other embeddings}
\label{sec:appB}

It seems a reasonable question whether the construction described in section~\ref{sec:fusion-ATL} is the only way to embed the small algebras
$\ATL{N_1}$ and $\ATL{N_2}$ into the big one $\ATL{N=N_1+N_2}$. If another embedding could be found, it could conceivably lead to different
branching rules and hence, by the Frobenius reciprocity, potentially to other fusion channels. We have made several attempts to find other
embeddings and here we give a brief account of one of the ideas that we have tried out.

The existence of an embedding hinges on the existence of  embeddings of two subalgebra translation operators $u^{(1)}$ and $u^{(2)}$. One of these embeddings was defined
in (\ref{u1-def})--(\ref{u2-def}). We recall that those are products of the full translation operator $u$ and braidings of one subsystem with
respect to the other. As previously pointed out, to ensure the crucial commutation $[u^{(1)},u^{(2)}] = 0$ it is necessary that in $u^{(1)}$ the
first subsystem braids completely over the second, and that in $u^{(2)}$ the second subsystem braids completely under the first---up to a
possible global swap of over and underpasses; see (\ref{utilde1-def})--(\ref{utilde2-def}). In particular, the braiding of one subsystem with
respect to the other cannot be a mixture using both over and underpasses, since then inevitably some strands will ``get stuck'' in the computation
of $u^{(1)} u^{(2)} - u^{(2)} u^{(1)}$.

We conclude from these observations that the only possible alternative is a modification of $u$ itself,
which corresponds to finding a non-trivial automorphism on the affine TL algebra. So rather than letting $u$ be a simple cyclic
shift of the strands, as shown in (\ref{u1-def})--(\ref{u2-def}), it appears natural to look for an alternative shift operator (henceforth denoted $\widetilde{u}$
to distinguish it from the usual $u$) that involves also some braiding.
We therefore try the following Ansatz (shown here for $N=4$):
\begin{equation}
\widetilde{u} \ = \ \begin{tikzpicture}[scale=0.6,baseline={([yshift=-3pt]current bounding box.east)}]
 \draw[thick] (1,0) -- (1,0.7) arc(180:90:0.3) -- (5,1);
 \draw[thick] (2,0) -- (2,1.7) arc(180:90:0.3) -- (5,2);
 \draw[fill,white] (2,1) circle(0.3);
 \draw (2,1) node {\scriptsize $L$};
 \draw[thick] (3,0) -- (3,2.7) arc(180:90:0.3) -- (5,3);
 \draw[fill,white] (3,1) circle(0.3);
 \draw (3,1) node {\scriptsize $K$};
 \draw[fill,white] (3,2) circle(0.3);
 \draw (3,2) node {\scriptsize $I$};
 \draw[thick] (4,0) -- (4,3.7) arc(180:90:0.3) -- (5,4);
 \draw[fill,white] (4,1) circle(0.3);
 \draw (4,1) node {\scriptsize $J$};
 \draw[fill,white] (4,2) circle(0.3);
 \draw (4,2) node {\scriptsize $H$};
 \draw[fill,white] (4,3) circle(0.3);
 \draw (4,3) node {\scriptsize $G$};
 \draw[thick] (0,1) -- (0.7,1) arc(-90:0:0.3) -- (1,4.7) arc(180:90:0.3) -- (1.7,5) arc(-90:0:0.3) -- (2,6);
 \draw[thick] (0,2) -- (1.7,2) arc(-90:0:0.3) -- (2,4.7) arc(180:90:0.3) -- (2.7,5) arc(-90:0:0.3) -- (3,6);
 \draw[fill,white] (1,2) circle(0.3);
 \draw (1,2) node {\scriptsize $F$};
 \draw[thick] (0,3) -- (2.7,3) arc(-90:0:0.3) -- (3,4.7) arc(180:90:0.3) -- (3.7,5) arc(-90:0:0.3) -- (4,6);
 \draw[fill,white] (1,3) circle(0.3);
 \draw (1,3) node {\scriptsize $E$};
 \draw[fill,white] (2,3) circle(0.3);
 \draw (2,3) node {\scriptsize $D$};
 \draw[thick] (0,4) -- (3.7,4) arc(-90:0:0.3) -- (4,4.7) arc(180:90:0.3) -- (5,5);
 \draw[fill,white] (1,4) circle(0.3);
 \draw (1,4) node {\scriptsize $C$};
 \draw[fill,white] (2,4) circle(0.3);
 \draw (2,4) node {\scriptsize $B$};
 \draw[fill,white] (3,4) circle(0.3);
 \draw (3,4) node {\scriptsize $A$};
 \draw[thick] (0,5) -- (0.7,5) arc(-90:0:0.3) -- (1,6);
\end{tikzpicture}
\qquad \qquad
\widetilde{u}^{-1} \ = \ \begin{tikzpicture}[xscale=0.6,yscale=-0.6,baseline={([yshift=-3pt]current bounding box.east)}]
 \draw[thick] (1,0) -- (1,0.7) arc(180:90:0.3) -- (5,1);
 \draw[thick] (2,0) -- (2,1.7) arc(180:90:0.3) -- (5,2);
 \draw[fill,white] (2,1) circle(0.3);
 \draw (2,1) node {\scriptsize $\bar{L}$};
 \draw[thick] (3,0) -- (3,2.7) arc(180:90:0.3) -- (5,3);
 \draw[fill,white] (3,1) circle(0.3);
 \draw (3,1) node {\scriptsize $\bar{K}$};
 \draw[fill,white] (3,2) circle(0.3);
 \draw (3,2) node {\scriptsize $\bar{I}$};
 \draw[thick] (4,0) -- (4,3.7) arc(180:90:0.3) -- (5,4);
 \draw[fill,white] (4,1) circle(0.3);
 \draw (4,1) node {\scriptsize $\bar{J}$};
 \draw[fill,white] (4,2) circle(0.3);
 \draw (4,2) node {\scriptsize $\bar{H}$};
 \draw[fill,white] (4,3) circle(0.3);
 \draw (4,3) node {\scriptsize $\bar{G}$};
 \draw[thick] (0,1) -- (0.7,1) arc(-90:0:0.3) -- (1,4.7) arc(180:90:0.3) -- (1.7,5) arc(-90:0:0.3) -- (2,6);
 \draw[thick] (0,2) -- (1.7,2) arc(-90:0:0.3) -- (2,4.7) arc(180:90:0.3) -- (2.7,5) arc(-90:0:0.3) -- (3,6);
 \draw[fill,white] (1,2) circle(0.3);
 \draw (1,2) node {\scriptsize $\bar{F}$};
 \draw[thick] (0,3) -- (2.7,3) arc(-90:0:0.3) -- (3,4.7) arc(180:90:0.3) -- (3.7,5) arc(-90:0:0.3) -- (4,6);
 \draw[fill,white] (1,3) circle(0.3);
 \draw (1,3) node {\scriptsize $\bar{E}$};
 \draw[fill,white] (2,3) circle(0.3);
 \draw (2,3) node {\scriptsize $\bar{D}$};
 \draw[thick] (0,4) -- (3.7,4) arc(-90:0:0.3) -- (4,4.7) arc(180:90:0.3) -- (5,5);
 \draw[fill,white] (1,4) circle(0.3);
 \draw (1,4) node {\scriptsize $\bar{C}$};
 \draw[fill,white] (2,4) circle(0.3);
 \draw (2,4) node {\scriptsize $\bar{B}$};
 \draw[fill,white] (3,4) circle(0.3);
 \draw (3,4) node {\scriptsize $\bar{A}$};
 \draw[thick] (0,5) -- (0.7,5) arc(-90:0:0.3) -- (1,6);
\end{tikzpicture}
\label{proposed-u}
\end{equation}
In these pictures, each capital letter can represent any of the braid generators $g_i$ or $g_i^{-1}$, the precise choice
being dictated by the following considerations. For starters, we note that each letter with an overbar must represent
the inverse of the corresponding letter without a bar, in order for $\widetilde{u}^{-1}$ to be the inverse of $\widetilde{u}$. Note also that whatever
the choice of under and overpasses, each line goes straight through the intersections, so that $\widetilde{u}$ indeed shifts the $N$
strands cyclically to the right (up to the braidings). However, unlike what happens in the usual $u$, see~(\ref{u1-def}), each strand
is not just shifted one position towards the right but makes a full extra tour, allowing for some extra degrees of freedom
via the choice of braidings.

We must now impose the required shift relations $\widetilde{u} e_i \widetilde{u}^{-1} = e_{i+1}$, with the indices considered mod $N$.
To this end, consider first
\begin{equation}
\widetilde{u} e_1 \widetilde{u}^{-1} \ = \ \begin{tikzpicture}[scale=0.6,baseline={([yshift=-3pt]current bounding box.east)}]
 \draw[thick] (1,0) -- (1,0.7) arc(180:90:0.3) -- (5,1);
 \draw[thick] (2,0) -- (2,1.7) arc(180:90:0.3) -- (5,2);
 \draw[fill,white] (2,1) circle(0.3);
 \draw (2,1) node {\scriptsize $L$};
 \draw[thick] (3,0) -- (3,2.7) arc(180:90:0.3) -- (5,3);
 \draw[fill,white] (3,1) circle(0.3);
 \draw (3,1) node {\scriptsize $K$};
 \draw[fill,white] (3,2) circle(0.3);
 \draw (3,2) node {\scriptsize $I$};
 \draw[thick] (4,0) -- (4,3.7) arc(180:90:0.3) -- (5,4);
 \draw[fill,white] (4,1) circle(0.3);
 \draw (4,1) node {\scriptsize $J$};
 \draw[fill,white] (4,2) circle(0.3);
 \draw (4,2) node {\scriptsize $H$};
 \draw[fill,white] (4,3) circle(0.3);
 \draw (4,3) node {\scriptsize $G$};
 \draw[thick] (0,1) -- (0.7,1) arc(-90:0:0.3) -- (1,4.7) arc(180:90:0.3) -- (1.7,5) arc(-90:0:0.3) -- (2,6);
 \draw[thick] (0,2) -- (1.7,2) arc(-90:0:0.3) -- (2,4.7) arc(180:90:0.3) -- (2.7,5) arc(-90:0:0.3) -- (3,6);
 \draw[fill,white] (1,2) circle(0.3);
 \draw (1,2) node {\scriptsize $F$};
 \draw[thick] (0,3) -- (2.7,3) arc(-90:0:0.3) -- (3,4.7) arc(180:90:0.3) -- (3.7,5) arc(-90:0:0.3) -- (4,6);
 \draw[fill,white] (1,3) circle(0.3);
 \draw (1,3) node {\scriptsize $E$};
 \draw[fill,white] (2,3) circle(0.3);
 \draw (2,3) node {\scriptsize $D$};
 \draw[thick] (0,4) -- (3.7,4) arc(-90:0:0.3) -- (4,4.7) arc(180:90:0.3) -- (5,5);
 \draw[fill,white] (1,4) circle(0.3);
 \draw (1,4) node {\scriptsize $C$};
 \draw[fill,white] (2,4) circle(0.3);
 \draw (2,4) node {\scriptsize $B$};
 \draw[fill,white] (3,4) circle(0.3);
 \draw (3,4) node {\scriptsize $A$};
 \draw[thick] (0,5) -- (0.7,5) arc(-90:0:0.3) -- (1,6);
 \draw[thick] (1,0) -- (1,-0.2) arc(180:270:0.3) -- (1.7,-0.5) arc(270:360:0.3) -- (2,0);
 \draw[thick] (1,-1.5) -- (1,-1.3) arc(180:90:0.3) -- (1.7,-1.0) arc(90:0:0.3) -- (2,-1.5);
 \draw[thick] (3,-1.5) -- (3,0);
 \draw[thick] (4,-1.5) -- (4,0);
\begin{scope}[yshift=-1.5cm,yscale=-1]
 \draw[thick] (1,0) -- (1,0.7) arc(180:90:0.3) -- (5,1);
 \draw[thick] (2,0) -- (2,1.7) arc(180:90:0.3) -- (5,2);
 \draw[fill,white] (2,1) circle(0.3);
 \draw (2,1) node {\scriptsize $\bar{L}$};
 \draw[thick] (3,0) -- (3,2.7) arc(180:90:0.3) -- (5,3);
 \draw[fill,white] (3,1) circle(0.3);
 \draw (3,1) node {\scriptsize $\bar{K}$};
 \draw[fill,white] (3,2) circle(0.3);
 \draw (3,2) node {\scriptsize $\bar{I}$};
 \draw[thick] (4,0) -- (4,3.7) arc(180:90:0.3) -- (5,4);
 \draw[fill,white] (4,1) circle(0.3);
 \draw (4,1) node {\scriptsize $\bar{J}$};
 \draw[fill,white] (4,2) circle(0.3);
 \draw (4,2) node {\scriptsize $\bar{H}$};
 \draw[fill,white] (4,3) circle(0.3);
 \draw (4,3) node {\scriptsize $\bar{G}$};
 \draw[thick] (0,1) -- (0.7,1) arc(-90:0:0.3) -- (1,4.7) arc(180:90:0.3) -- (1.7,5) arc(-90:0:0.3) -- (2,6);
 \draw[thick] (0,2) -- (1.7,2) arc(-90:0:0.3) -- (2,4.7) arc(180:90:0.3) -- (2.7,5) arc(-90:0:0.3) -- (3,6);
 \draw[fill,white] (1,2) circle(0.3);
 \draw (1,2) node {\scriptsize $\bar{F}$};
 \draw[thick] (0,3) -- (2.7,3) arc(-90:0:0.3) -- (3,4.7) arc(180:90:0.3) -- (3.7,5) arc(-90:0:0.3) -- (4,6);
 \draw[fill,white] (1,3) circle(0.3);
 \draw (1,3) node {\scriptsize $\bar{E}$};
 \draw[fill,white] (2,3) circle(0.3);
 \draw (2,3) node {\scriptsize $\bar{D}$};
 \draw[thick] (0,4) -- (3.7,4) arc(-90:0:0.3) -- (4,4.7) arc(180:90:0.3) -- (5,5);
 \draw[fill,white] (1,4) circle(0.3);
 \draw (1,4) node {\scriptsize $\bar{C}$};
 \draw[fill,white] (2,4) circle(0.3);
 \draw (2,4) node {\scriptsize $\bar{B}$};
 \draw[fill,white] (3,4) circle(0.3);
 \draw (3,4) node {\scriptsize $\bar{A}$};
 \draw[thick] (0,5) -- (0.7,5) arc(-90:0:0.3) -- (1,6);
\end{scope}
\end{tikzpicture}
\end{equation}
The twists at letters $L$ and $\bar{L}$ give rise to opposite factors and can hence be undone. The resulting ``tongue'' can be pulled across
strands 3 and 4 on condition that $I = K$ and $H = J$. Any other choice means that the tongue gets entangled with those strands, which is
unacceptable since this would make impossible further reductions (we must obtain $e_2$ in the end). With those constraints the situation is
now:
\begin{equation}
\widetilde{u} e_1 \widetilde{u}^{-1} \ = \ \begin{tikzpicture}[scale=0.6,baseline={([yshift=-3pt]current bounding box.east)}]
 \draw[thick] (3,0) -- (3,2.7) arc(180:90:0.3) -- (5,3);
 \draw[thick] (4,0) -- (4,3.7) arc(180:90:0.3) -- (5,4);
 \draw[fill,white] (4,3) circle(0.3);
 \draw (4,3) node {\scriptsize $G$};
 \draw[thick] (0.5,1) -- (0.7,1) arc(-90:0:0.3) -- (1,4.7) arc(180:90:0.3) -- (1.7,5) arc(-90:0:0.3) -- (2,6);
 \draw[thick] (0.5,2) -- (1.7,2) arc(-90:0:0.3) -- (2,4.7) arc(180:90:0.3) -- (2.7,5) arc(-90:0:0.3) -- (3,6);
 \draw[fill,white] (1,2) circle(0.3);
 \draw (1,2) node {\scriptsize $F$};
 \draw[thick] (0,3) -- (2.7,3) arc(-90:0:0.3) -- (3,4.7) arc(180:90:0.3) -- (3.7,5) arc(-90:0:0.3) -- (4,6);
 \draw[fill,white] (1,3) circle(0.3);
 \draw (1,3) node {\scriptsize $E$};
 \draw[fill,white] (2,3) circle(0.3);
 \draw (2,3) node {\scriptsize $D$};
 \draw[thick] (0,4) -- (3.7,4) arc(-90:0:0.3) -- (4,4.7) arc(180:90:0.3) -- (5,5);
 \draw[fill,white] (1,4) circle(0.3);
 \draw (1,4) node {\scriptsize $C$};
 \draw[fill,white] (2,4) circle(0.3);
 \draw (2,4) node {\scriptsize $B$};
 \draw[fill,white] (3,4) circle(0.3);
 \draw (3,4) node {\scriptsize $A$};
 \draw[thick] (0,5) -- (0.7,5) arc(-90:0:0.3) -- (1,6);
 \draw[thick] (0.5,1) -- (0.3,1) arc(270:180:0.3) -- (0,1.7) arc(180:90:0.3) -- (0.5,2);
\begin{scope}[yshift=1.0cm,yscale=-1]
 \draw[thick] (3,0) -- (3,2.7) arc(180:90:0.3) -- (5,3);
 \draw[thick] (4,0) -- (4,3.7) arc(180:90:0.3) -- (5,4);
 \draw[fill,white] (4,3) circle(0.3);
 \draw (4,3) node {\scriptsize $\bar{G}$};
 \draw[thick] (0.5,1) -- (0.7,1) arc(-90:0:0.3) -- (1,4.7) arc(180:90:0.3) -- (1.7,5) arc(-90:0:0.3) -- (2,6);
 \draw[thick] (0.5,2) -- (1.7,2) arc(-90:0:0.3) -- (2,4.7) arc(180:90:0.3) -- (2.7,5) arc(-90:0:0.3) -- (3,6);
 \draw[fill,white] (1,2) circle(0.3);
 \draw (1,2) node {\scriptsize $\bar{F}$};
 \draw[thick] (0,3) -- (2.7,3) arc(-90:0:0.3) -- (3,4.7) arc(180:90:0.3) -- (3.7,5) arc(-90:0:0.3) -- (4,6);
 \draw[fill,white] (1,3) circle(0.3);
 \draw (1,3) node {\scriptsize $\bar{E}$};
 \draw[fill,white] (2,3) circle(0.3);
 \draw (2,3) node {\scriptsize $\bar{D}$};
 \draw[thick] (0,4) -- (3.7,4) arc(-90:0:0.3) -- (4,4.7) arc(180:90:0.3) -- (5,5);
 \draw[fill,white] (1,4) circle(0.3);
 \draw (1,4) node {\scriptsize $\bar{C}$};
 \draw[fill,white] (2,4) circle(0.3);
 \draw (2,4) node {\scriptsize $\bar{B}$};
 \draw[fill,white] (3,4) circle(0.3);
 \draw (3,4) node {\scriptsize $\bar{A}$};
 \draw[thick] (0,5) -- (0.7,5) arc(-90:0:0.3) -- (1,6);
 \draw[thick] (0.5,1) -- (0.3,1) arc(270:180:0.3) -- (0,1.7) arc(180:90:0.3) -- (0.5,2);
\end{scope}
\end{tikzpicture}
\end{equation}
The tongue is now situated next to the letter $F$ and we can repeat the preceding argument. The twists at $F$ and $\bar{F}$ are opposite
and can be undone. Further, the tongue can be retracted across the two horizontal strands provided that $D=E$ and $B=C$. This gives
then
\begin{equation}
\widetilde{u} e_1 \widetilde{u}^{-1} \ = \ \begin{tikzpicture}[scale=0.6,baseline={([yshift=-3pt]current bounding box.east)}]
 \draw[thick] (3,2) -- (3,2.7) arc(180:90:0.3) -- (5,3);
 \draw[thick] (4,2) -- (4,3.7) arc(180:90:0.3) -- (5,4);
 \draw[fill,white] (4,3) circle(0.3);
 \draw (4,3) node {\scriptsize $G$};
 \draw[thick] (2,6) -- (2,5.3) arc(-180:-90:0.3) -- (2.7,5) arc(-90:0:0.3) -- (3,6);
 \draw[thick] (0,3) -- (2.7,3) arc(-90:0:0.3) -- (3,4.7) arc(180:90:0.3) -- (3.7,5) arc(-90:0:0.3) -- (4,6);
 \draw[thick] (0,4) -- (3.7,4) arc(-90:0:0.3) -- (4,4.7) arc(180:90:0.3) -- (5,5);
 \draw[fill,white] (3,4) circle(0.3);
 \draw (3,4) node {\scriptsize $A$};
 \draw[thick] (0,5) -- (0.7,5) arc(-90:0:0.3) -- (1,6);
\begin{scope}[yshift=4.5cm,yscale=-1]
 \draw[thick] (3,2) -- (3,2.7) arc(180:90:0.3) -- (5,3);
 \draw[thick] (4,2) -- (4,3.7) arc(180:90:0.3) -- (5,4);
 \draw[fill,white] (4,3) circle(0.3);
 \draw (4,3) node {\scriptsize $\bar{G}$};
 \draw[thick] (2,6) -- (2,5.3) arc(-180:-90:0.3) -- (2.7,5) arc(-90:0:0.3) -- (3,6);
 \draw[thick] (0,3) -- (2.7,3) arc(-90:0:0.3) -- (3,4.7) arc(180:90:0.3) -- (3.7,5) arc(-90:0:0.3) -- (4,6);
 \draw[thick] (0,4) -- (3.7,4) arc(-90:0:0.3) -- (4,4.7) arc(180:90:0.3) -- (5,5);
 \draw[fill,white] (3,4) circle(0.3);
 \draw (3,4) node {\scriptsize $\bar{A}$};
 \draw[thick] (0,5) -- (0.7,5) arc(-90:0:0.3) -- (1,6);
\end{scope}
\end{tikzpicture}
\end{equation}
In the final step we pull the tongue at letters $G$ and $\bar{G}$ to the right across strand 4, via the periodic boundary condition, and notice
that the opposite braidings at $A$ and $\bar{A}$ can hence be undone. The result is $\widetilde{u} e_1 \widetilde{u}^{-1} = e_2$, as desired.

We next consider $\widetilde{u} e_2 \widetilde{u}^{-1}$ in exactly the same way. This leads to the constraints $K = L$, $G = H$, $E = F$ and $A = B$, and in the end
we obtain $e_3$ as we should. Similar considerations on $\widetilde{u} e_3 \widetilde{u}^{-1}$ lead to $J = K$, $H = I$, $C = E$ and $B = D$, and we obtain
$e_0$ in the end.

Summarising this far, we have
\begin{equation}
 A = B = C = D = E = F \qquad \mbox{and} \qquad G = H = I = J = K = L \,.
 \label{appB-constraints}
\end{equation}

The ultimate stage of the computation is to consider $\widetilde{u} e_0 \widetilde{u}^{-1}$. We find that this gives $e_1$, as it should, on condition that $A = G^{-1}$.
The only remaining degree of freedom is thus $A$. But redrawing (\ref{proposed-u}) with either $A = g_i$ or $A = g_i^{-1}$, it is easy to see that in
both cases
\begin{equation}
 \widetilde{u} = u^{N+1} \,,
\end{equation}
with $u$ the usual one-step shift operator. This is just a modification by the central element $u^N$.
 
 It remains to check the last of the defining relations \eqref{TLpdef-u2}, namely $\widetilde{u}^2 e_{N-1} = e_1 \dots e_{N-1}$. The diagrammatic interpretation of the
right-hand side is that there is an arc between the last two sites on the bottom and between the first two sites on the top, while the first $N-2$ strands on the bottom
shift two steps to the right so as to connect to the last $N-2$ strands on the top. This coincides indeed with $u^2 e_{N-1}$ in the usual case. However, an easy diagrammatic
computation shows that with $\widetilde{u}^2 e_{N-1}$ the first $N-2$ strands on the bottom shift two steps to the right and in addition make two full tours around the system.
Therefore the last relation \eqref{TLpdef-u2} fails in the algebra, and we have---in spite of our efforts---not obtained the desired automorphism.
Our other attempts of defining alternative embeddings have been equally fruitless.

\end{document}